\documentclass[a4paper, 11pt]{article}
\usepackage{LocalPackages}
\usepackage{authblk}

\begin{document}

\title{Error correcting codes and spatial coupling}
\date{}
\author[1]{Rafah El-Khatib}
\author[2]{Jean Barbier}
\author[3]{Ayaka Sakata}
\author[1]{R{\" u}diger Urbanke}
\affil[1]{EPFL, Lausanne, Switzerland}
\affil[2]{Ecole Normale Sup\'erieure, Paris, France}
\affil[3]{RIKEN, Wako, Japan}

\maketitle

These are notes from the lecture of R{\" u}diger Urbanke given at the autumn school
"Statistical Physics, Optimization, Inference, and Message-Passing Algorithms",
that took place in Les Houches, France from Monday September 30th, 2013, till
Friday October 11th, 2013. The school was organized by Florent Krzakala from
UPMC and ENS Paris, Federico Ricci-Tersenghi from La Sapienza Roma, Lenka
Zdeborov\'a from CEA Saclay and CNRS, and Riccardo Zecchina from Politecnico
Torino. 

\tableofcontents

\newpage

\section{Polar Codes}
\subsection{Motivation}
\begin{figure}
\centering
\includegraphics[scale=0.3]{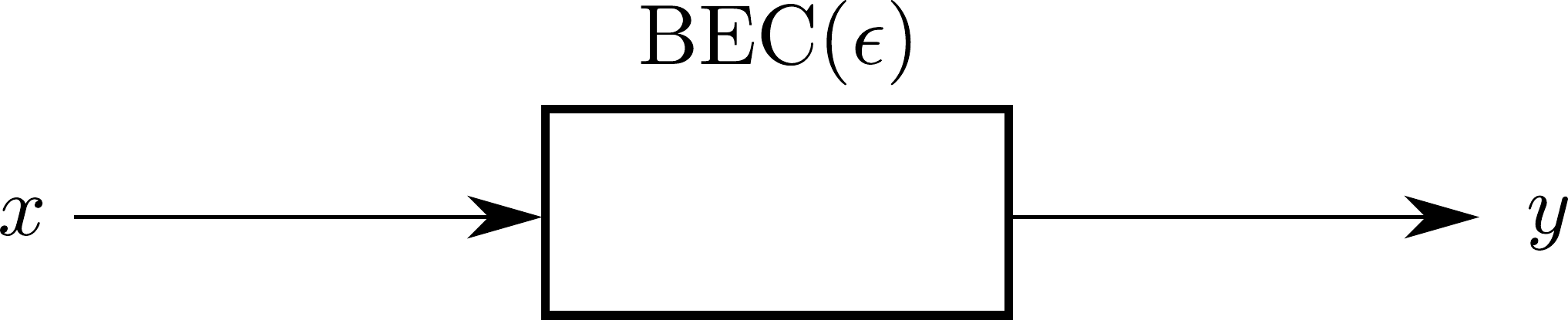}
\caption{A transmission scheme of scalar quantities over the binary erasure channel with parameter $\epsilon$.}
\label{fig:becScalar}
\end{figure}

Consider the transmission scheme depicted in Fig.~\ref{fig:becScalar}
where one bit $x\in\{0,1\}$ is sent over a channel that either erases $x$
with probability (w.p.) $\epsilon$ or passes the bit unchanged w.p. $1-\epsilon$. This channel is called the
Binary Erasure Channel with parameter $\epsilon$ and we will denote it as BEC($\epsilon$).
The receiver thus receives the symbol $y\in\{0, 1, ?\}$ where 
\begin{equation}\label{eqn:BEC}
y=
\begin{cases}
x & \text{ w.p. }1-\epsilon,
\\
? & \text{ w.p. }\epsilon.
\end{cases}
\end{equation}
We want to recover the transmitted bit at the receiver and for this
purpose the receiver forms an ``estimate'' of $x$ given the received
symbol $y$, denote this estimate as $\hat x(y)$; our goal is to
minimize the quantity Pr($\{\hat x(y) \neq x \}$), i.e., we want
to minimize the {\em probability of error}.

If we are only sending a single bit then we cannot hope to estimate
the transmitted bit reliably in case it was erased. There is simply
not enough information available. The picture changes if we are
sending a {\em block} of bits.

\begin{figure}
\centering
\includegraphics[scale=0.3]{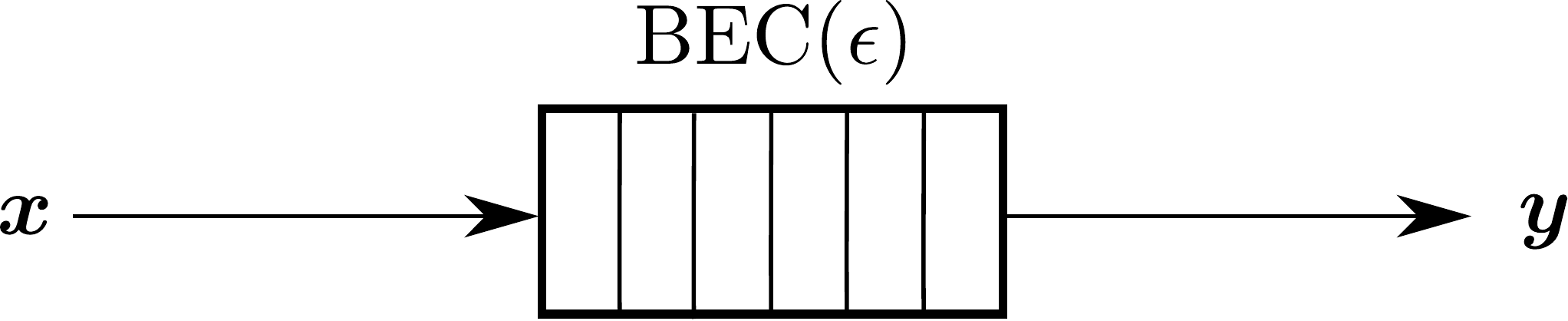}
\caption{A transmission scheme of vector quantities over the binary erasure channel with parameter $\epsilon$.}
\label{fig:becVector}
\end{figure}

Consider therefore the slightly more general setting shown in Fig.~\ref{fig:becVector} where a vector $\mathbf{x}\in\{0,1\}^n$ is sent
on the same channel, the BEC($\epsilon$). At the receiver, the
vector $\mathbf{y}\in\{0,1\}^n$ is received such that each component
of this vector follows the rules in \eqref{eqn:BEC}, i.e., each
component is erased independently from all other components with
probability $\epsilon$.  It is easy to determine the
expected number of erased and non-erased bits for this scenario, namely
\begin{align*}
\mathbb{E}[|\{y_i=? \}|]&=n\epsilon,\\
\mathbb{E}[|\{y_i\neq ? \}|]&=n(1-\epsilon).
\end{align*}
The standard deviation associated to these values is
$\sigma=\sqrt{n\epsilon(1-\epsilon)}$. Similar to before, we are
interested in determining the transmitted vector $\mathbf{x}$ given
the received vector $\mathbf{y}$, and for this purpose we form the
estimate $\hat{\mathbf{x}}(\mathbf{y})$, and, as before,
Pr($\{\hat{\mathbf{x}}(\mathbf{y}) \neq \mathbf{x} \}$) is the
probability of error. If this probability of error is ``small'' then
we say that we achieve a reliable transmission.

It is now natural to ask the following question: how many bits can
we reliably transmit over such a channel, measured as a function
of the vector length $n$, when $n\rightarrow +\infty$? We derive
first an upper bound and then a matching lower bound.

\subsection{Upper Bound} 
Assume that we are aided by a genie that tells the transmitter the positions
that will be erased ahead of time. More formally, let $S \subseteq
[n]$ be the set of erasures and assume that we know this set before
sending the vector $\mathbf{x}$.  Some thought then shows that the
optimal strategy consists of sending our information in the positions
$[n] \setminus S$ and to fill the positions $S$ with dummy bits.
The receiver then simply reads off the positions in $[n] \setminus
S$ to recover the transmitted information and this way estimate at
the receiver is perfect and we never make an error. This is also
clearly the maximum amount of information that we can transmit
reliably in this scenario.

Since $\mathbb{E}[|[n]\setminus S|]=n(1-\epsilon)$ it follows that
the fraction of channel uses on which information can be sent
reliably is equal to $1-\epsilon$. This fraction is called the
``transmission rate" and the highest possible rate for which reliable
transmission is possible is called the ``capacity.'' Since we have
just shown that the capacity of the genie-aided transmission is
$1-\epsilon$ it follows that the real capacity (i.e., the capacity
without the genie) is {\em at most} $1-\epsilon$.

\subsection{Lower Bound}
To prove that the capacity is equal to $1-\epsilon$ we now derive
a matching lower bound by describing a scheme which allows reliable
transmission all the way up to a rate of $1-\epsilon$.  The typical
way to prove this lower bound is by using a so-called ``random
coding" argument.  This argument proceed by showing that ``randomly''
chosen codes from a suitably defined ``ensemble'' of codes work
with high probability.\footnote{A {\em code} is a subset of the set
of all binary $n$-tuples and typically this subset is chosen in
such a way that the individual codewords are well separated. This
ensures that even with some of the components being erased, the
receiver can still figure out which of the codewords was sent.}
This argument has the advantage that it is relatively simple and
short. But on the downside, the argument is non-constructive and
in addition does not take the complexity of the scheme into account.

Instead, we will describe an explicit scheme which in addition is
also of low complexity.  It is called the ``polar coding'' scheme.
This scheme is fairly recent but the basic idea has already proven
to be fundamental in a variety of areas [E. Arikan, 2008].

Let us first make an observation. For the BEC$(\epsilon=0)$ the
capacity is $1-\epsilon=1$. That is, we can fill in the entire
vector $\mathbf{x}$ with information bits and recover them reliably.
If, on the other hand, the erasure probability is $\epsilon=1$,
then the capacity is $0$ and there is no use of sending any information
in $\mathbf{x}$. Both of these cases are thus {\em easy} to deal
with in the sense that we know what to do.  This observation extends
to cases where $\epsilon \sim 0$ and $\epsilon\sim 1$.  More
precisely, we lose very little in the case when $\epsilon\sim 1$
by not using the channel, and if $\epsilon$ is very small (compared
to $n$) then we can still use all components of the block and most
of the time the whole block will arrive without erasures.  Only
once in a while will we not be able to recover the block, and this
simply results in a small probability of error.

Let us now introduce the basic idea of {\em polarization}. Consider
the transmission scheme in Fig.~\ref{fig:polar2x2}. Two bits,
call them $U_1$ and $U_2$, are chosen uniformly at random from
$\{0,1\}$ and are encoded into two other bits, denoted by $X_1$ and
$X_2 \in\{0,1\}$, as follows (all operations are over the binary
field):
\begin{figure}
\centering
\includegraphics{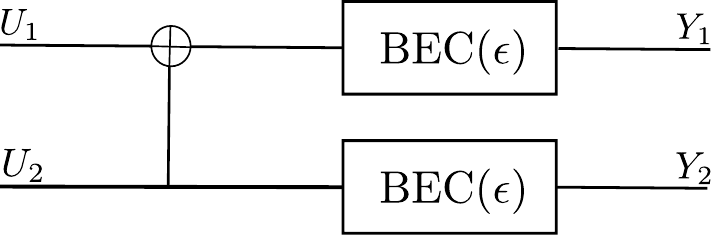}
\caption{A one-step polarization transformation for the BEC($\epsilon$)'s.}
\label{fig:polar2x2}
\end{figure}

\begin{align}
X_1&=U_1+U_2, \label{equ:x1}\\
X_2&=U_2. \label{equ:x2}
\end{align}
Equivalently, we can describe this relationship in matrix form,
\begin{equation*}
\begin{bmatrix}
U_1 & U_2
\end{bmatrix}\quad
\begin{bmatrix}
1 & 0\\
1 & 1
\end{bmatrix}=
\begin{bmatrix}
X_1 & X_2
\end{bmatrix}.
\end{equation*}
Assume that we receive $\mathbf{Y}=(Y_1\;Y_2)$ and that we want to
estimate $U_1$ given $\mathbf{Y}=(Y_1\;Y_2)$, where $U_2$ is unknown
(and has a uniform prior).  We denote by $\hat U_1(\mathbf{Y})$
this estimate.  Note that from (\ref{equ:x1}) we know that $U_1=X_1+U_2$
(there are no signs in the binary field).  If we combine this with
(\ref{equ:x2}) we see that $U_1=X_1+X_2$. Note further that $Y_1$
and $Y_2$ are the result of transmitting $X_1$ and $X_2$, respectively,
through independent erasure channels.

We therefore see that we can reconstruct $U_1$ if and only if neither
$Y_1$ nor $Y_2$ are erasures, so that $X_1=Y_1$ and
$X_2=Y_2$.  Hence, we have
\begin{align*}
\hat U_1(\mathbf{Y})=
\begin{cases}
Y_1+Y_2,  & \text{ if } Y_1\neq ?\wedge Y_2\neq ?,
\\
?, & \text{ otherwise.}
\end{cases}
\end{align*}
Note that the bits $X_1$ and $X_2$ are sent over independent erasure channels and that
\begin{align}
Pr(\{Y_1=?\})&=\epsilon,\label{eqn:Yerasure}\\
Pr(\{Y_2=?\})&=\epsilon.
\end{align}
Therefore,
\begin{align*}
Pr(\{\hat U_1(\mathbf{Y})=U_1\})&=Pr(\{Y_1\neq ? \wedge Y_2\neq ? \})=(1-\epsilon)^2,\\
Pr(\{\hat U_1(\mathbf{Y})=?\})&=1-(1-\epsilon)^2=\epsilon(2-\epsilon)>\epsilon.
\end{align*}
As we can see, the probability that $U_1$ is erased is strictly
larger than $\epsilon$ (unless $\epsilon=1$). So this does not seem
to be a very good scheme.
\index{Y} Why then would we use this transform,
which is called the {\em polar} transform?  As we will see shortly,
estimating the bit $U_2$ is in fact easier than the original problem,
and estimating the bit $U_1$ is more difficult as we just discussed.
The key point is that both of these tasks are closer to the two
trivial scenarios ($\epsilon=0$ and $\epsilon=1$) and by recursing
this transform we will be able to approach these trivial cases
closer and closer. Once we are sufficiently close no extra coding
will be necessary since we know how to deal with these two cases.

\begin{figure}
\centering
\includegraphics{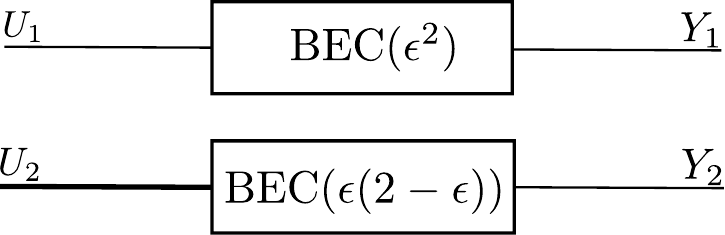}
\caption{A transmission scheme equivalent to the one-polarization step of two BEC($\epsilon$)'s.}
\label{fig:indep2x2}
\end{figure}

Let us now look at the problem of estimating $U_2$.  For this task,
we will assume that a genie tells us the true value of $U_1$. We
will soon see that in fact we have this information at the receiver
as long as we decode the various bits in the appropriate order.
Therefore, this assumption is in fact realistic. Let us summarize,
we want to estimate $U_2$ given $U_1$ and $\mathbf{Y}$. Reconsider
our two basic equations.  First, rewrite (\ref{equ:x1}) as $U_2=X_1+U_1$
and note that by assumption $U_1$ is known.  Further, write
(\ref{equ:x2}) as $U_2=X_2$. We therefore see that we have two
estimates of $U_2$ available at the receiver and that these two
estimates are conditionally independent since $X_1$ and $X_2$ are
transmitted over two independent channels (and $U_1$ is a known
constant). We conclude that we will be able to recover $U_2$ as
long as at least one of $Y_1$ and $Y_2$ are not erased. Let us summarize,
we have
\begin{align*}
\hat{U}_2(U_1, \mathbf{Y})=
\begin{cases}
Y_2, & \text{ if }Y_2 \neq ?,
\\
Y_1+U_1, & \text{ if } Y_2 =?\wedge Y_1\neq ?,
\\
?, & \text{ otherwise,}
\end{cases}
\end{align*}
and
\begin{align}
Pr(\{\hat U_2(U_1,\mathbf{Y})=? \})&=\epsilon^2 < \epsilon.
\end{align} 
Assume now that we estimate $U_1$ and $U_2$ successively using the following estimators.
\begin{align*}
\hat U_1&=\hat U_1 (\mathbf{Y}),\\
\hat U_2&=
\begin{cases}
\hat U_2 (\hat U_1,\mathbf{Y}) & \text{ if }U_1\neq ?,
\\
? & \text{ otherwise.}
\end{cases}
\end{align*}
Then, 
\begin{align}
Pr(\{\hat U_1(\mathbf{Y})=? \vee  \hat U_2(\hat U_1 (\mathbf{Y}),\mathbf{Y})=? \})&=Pr(\{\hat U_1(\mathbf{Y})=? \vee  \hat U_2(U_1,\mathbf{Y})=? \})\label{eqn:PolarEquiv}\\
&\leq Pr(\{\hat U_1(\mathbf{Y})=?\}) +Pr(\{\hat U_2(U_1,\mathbf{Y})=? \})\nonumber \\
& = \epsilon (2-\epsilon) + \epsilon^2 = 2 \epsilon.
\end{align} 
This has the following interpretation. In terms of this union bound, the successive decoder is as good as the scenario shown in Fig.~\ref{fig:indep2x2}, where we have two independent BEC's with different parameters.
The scheme in Fig.~\ref{fig:polar2x2} is obtained
from that in Fig.~\ref{fig:becScalar} using the following
relationship. Given two BEC's with parameter $\epsilon$ each, we
obtain a BEC with parameter $\epsilon(2-\epsilon)$, which is called
the ``-- channel" and a BEC with parameter $\epsilon^2$, which is
called the ``+ channel". Note further that the sum of the capacities
of these two channels is $1-\epsilon(2-\epsilon) +
1-\epsilon^2=2(1-\epsilon)$.  In other words, the average capacity
of these two channels is equal to the original capacity. So we have
lost nothing in terms of capacity by using this transform with the
particular successive decoding algorithm. If we look at the transform
itself this is not to surprising. After all, this transform is
invertible, and hence lossless.

Consider the scheme in Fig.~\ref{fig:polar2x2}
with $\epsilon=0.5$. Then the equivalent scheme in Fig.~\ref{fig:indep2x2} consists of two cascaded channels BEC($0.75$)
and BEC($0.25$). Notice that the average erasure probability over
the two channels is $\frac{0.75+0.25}{2}=\epsilon$.

This procedure of starting with two independent channels, combining
them, and then separating them again into two channels constitutes
one ``polarization step".

\begin{figure}
\centering
\includegraphics{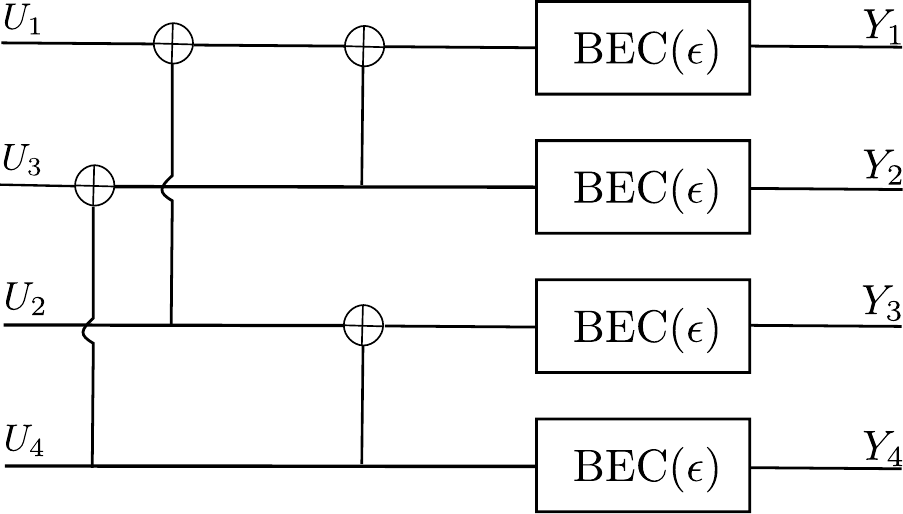}
\caption{A two-polarization step of four BEC($\epsilon$)'s.}
\label{fig:polar4x4}
\end{figure}

Rather than performing only a single step, we can now recurse. Let
us look explicitly at one further step, as shown in Fig.~\ref{fig:polar4x4}. Note that in the second step we combine ``like"
channels and that we decode successively in a very particular order,
namely $U_1$, $U_2$, $U_3$, $U_4$.
The erasure probabilities that we get for the four resulting ``synthetic'' channels are as follows:
\begin{itemize}
\item $U_1$ ``sees" the BEC w.p. $\delta(2-\delta)$ where $\delta=\epsilon(2-\epsilon)$.
\item $U_2$ ``sees" the BEC w.p. $\delta^2$ where $\delta=\epsilon(2-\epsilon)$.
\item $U_3$ ``sees" the BEC w.p. $\delta(2-\delta)$ where $\delta=\epsilon^2$.
\item $U_4$ ``sees" the BEC w.p. $\delta^2$ where $\delta=\epsilon^2$.
\end{itemize}

\begin{figure}
\centering
\includegraphics{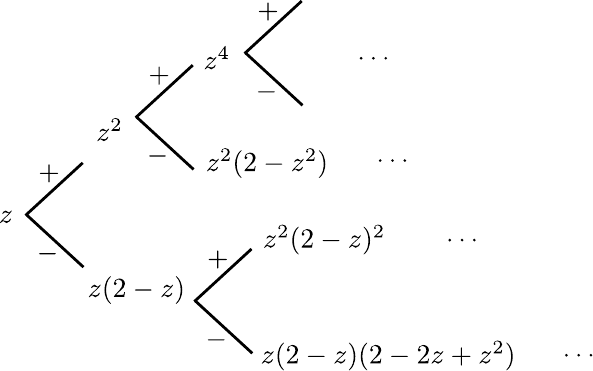}
\caption{A tree diagram that tracks the erasure probabilities obtained by polarization.}
\label{fig:zDiagram}
\end{figure}

Clearly, we can recurse this procedure $n\in\mathbb{N}$ times to
create from $2^n=N$ independent channels with parameter $\epsilon$, $N$ ``new"
(sometimes called synthetic) channels with parameters $\epsilon_i$,
$i\in 0\dots N-1$. The parameters evolve at each polarization step
according to the rules
\begin{align}
z&\rightarrow z(2-z),\label{eqn:zTransform}\\
z&\rightarrow z^2.\nonumber 
\end{align}
The evolution of the erasure probabilities upon this recursion can
be seen as an expansion of the tree diagram in Fig.~\ref{fig:zDiagram}.
Notice that the mean of each column, with respect to the uniform
distribution, is constant, namely equal to $\epsilon$. This is true
since
\begin{align*}
\frac{z^2+z(2-z)}{2}=z.
\end{align*}
This implies that the overall capacity stays preserved.

Recall now the motivation for using this transform. We know how to
deal with trivial and perfect channels and we hope that by applying
a sufficient number of these transforms the resulting synthetic
channels will all either become trivial or perfect. If this is
indeed the case, and since we know that the overall capacity is preserved,
it must be true that the proportion of perfect channels is equal
to the capacity of the original channel. Therefore, if we send our
bits over the perfect channels and fix the trivial channels to some
known value we will be able to transmit reliably arbitrarily close
to capacity.  It remains to show that this ``polarization'' of the
channels towards these extreme points is indeed the case.

\begin{figure}
\centering
\includegraphics{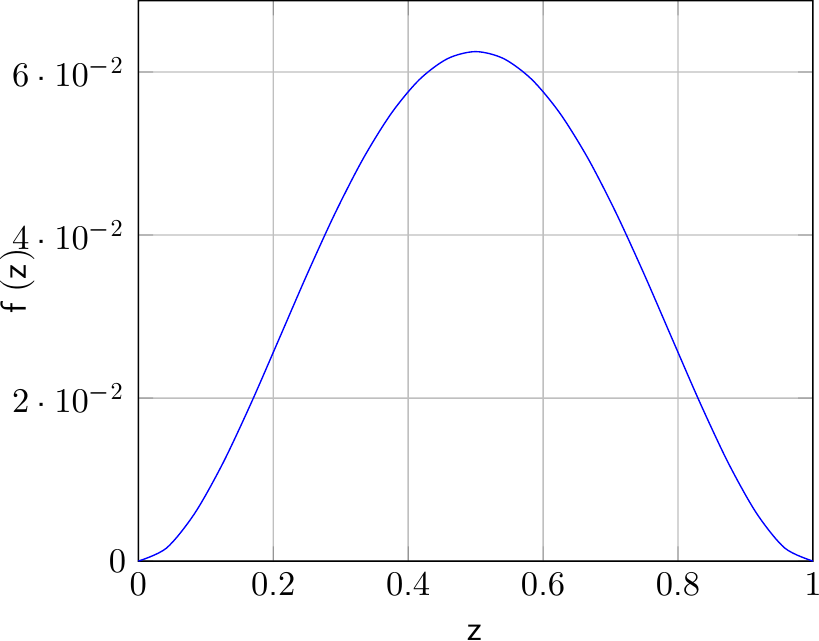}
\caption{The plot of $f(z)$.}
\label{fig:plotFz}
\end{figure}

\begin{figure}
\centering
\includegraphics{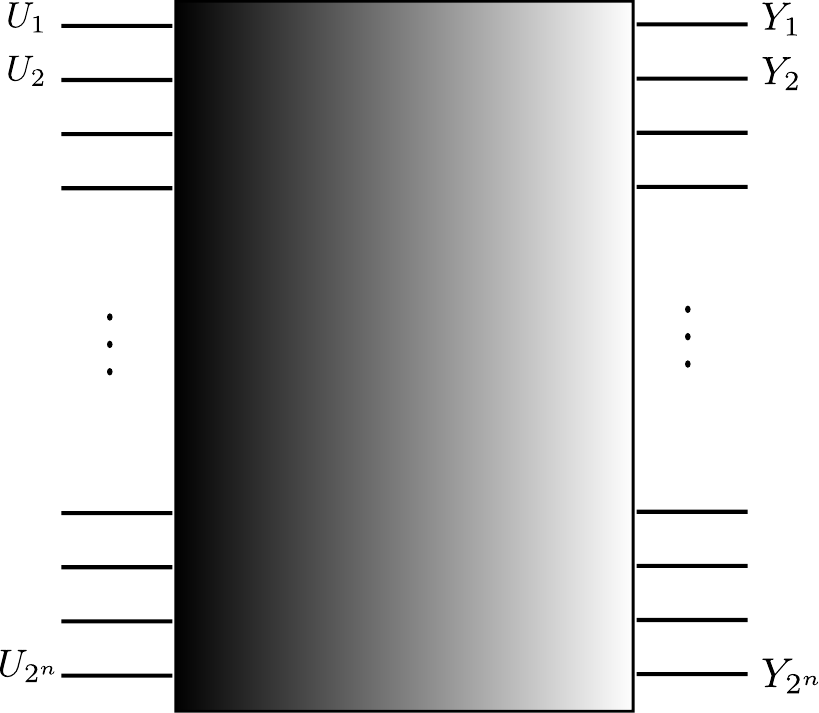}
\caption{A polarization scheme of $n$ steps.}
\label{fig:blackBox}
\end{figure}

Towards this goal, let us look at the second moment associated to this transformation $\rho_n^n$,
\begin{align*}
\rho_n^2=\frac{z^4+z^2(2-z)^2}{2}=z^4+2z^2(1-z).
\end{align*}
Consider $f(z)=z^4+2z^2(1-z)-z^2=z^2(z^2-2z+1)=z^2(z-1)^2$ which
represents the difference of the second moment after the transform
and before the transform.  Fig.~\ref{fig:plotFz} shows the plot
of $f(z)$.  Note that $f(z)>0$, $z\in(0,1)$, and that $f(0)=f(1)=0$. This
means the following: Consider the $n^{th}$ column and let $\mu_n$
and $\rho_n^2$ denote the mean and second moment, respectively. We
have seen that for $n\in\{0,1,\dots\}=\mathbb{N}$,
\begin{align*}
\mu_n&=\epsilon,\\
\rho_n^2& \;\mathrm{ is}\;\mathrm{ increasing}\;\mathrm{and} \;\rho_n^2\leq 1.
\end{align*}
Thus, $\lim\limits_{n\rightarrow +\infty}\rho_n^2=\rho_\infty^2$
exists. Note further that as long as a non-zero probability mass
lies strictly bounded away from $0$ and $1$, then the increase in
the second moment is strict. It is therefore clear that the limiting
distribution must be the one where all the mass is located either at
$0$ or at $1$. This is made precise in the following statement.
Let $z_{nj}$, $j\in\{1,2,\dots,2^n\}$, denote the $2^n$ numbers in
the $n^{th}$ column. For $\delta\in[0,\frac{1}{2}]$, define
\begin{equation*}
S_{n}(\delta)=\{j:\;\delta\leq z_{nj}\leq 1-\delta \}.
\end{equation*}
Then for any $\delta>0$, $\lim\limits_{n\rightarrow
+\infty}|S_n(\delta)|2^{-n}=0$. In words, all but a sublinear
fraction of channels is either ``good" or ``bad".
For a fixed $\delta>0$, we call a channel ``good" if it belongs to
the set $G_n(\delta)=\{j:\;|z_{nj}|<\delta \}$, and ``bad" if it
belongs to the set $B_n(\delta)=\{j:\;|1-z_{nj}|<\delta \}$.
Fix $\delta \in [0, \frac12]$.
Since $\lim\limits_{n\rightarrow +\infty}\rho_n^2=\rho_\infty^2$
exists, then for all $\Delta>0$, there exists $n_0\in\mathbb{N}$
so that
\begin{equation*}
\rho_n^2>\rho_\infty^2-\Delta \min\{f(\delta),f(1-\delta) \}
\end{equation*}
for all $n\geq n_0$. We claim that for all $n\geq n_0$,
$|S_n(\delta)|2^{-n}\leq \Delta$. Since this is true for all
$\Delta>0$, the claim will follow.
Assume that $|S_n(\delta)|2^{-n}>\Delta$. This means that there are
at least $\Delta 2^n$ numbers $z_{nj}$ in the range $[\delta,1-\delta]$.
It follows that the second moment must go up in the next iteration
by at least $\Delta \min\{f(\delta),f(1-\delta) \}$. But this would
imply that $\rho_{n+1}^2>\rho_\infty^2$, which is a contradiction.
A more careful analysis shows that, for $0\leq\beta<\frac{1}{2}$, with $N=2^n$, we have 
\begin{align*}
\lim\limits_{n\rightarrow +\infty}|S_n(2^{-2^{\beta n}})|2^{-n}&=0,\\
\lim\limits_{n\rightarrow +\infty}|G_n(2^{-2^{\beta n}})|2^{-n}&=1-\epsilon,\\
\lim\limits_{n\rightarrow +\infty}|B_n(2^{-2^{\beta n}})|2^{-n}&=\epsilon.
\end{align*}
This gives rise to the scheme shown in Fig.~\ref{fig:blackBox}.
Consider a polar code that is polarized $n$ times, where $n$ is
chosen to be ``sufficiently" large. The code thus has $2^n$ channels,
input bits $U_1,\dots,U_{2^n}$ and output bits $Y_1,\dots,Y_{2^n}$.
``Freeze" the channels $j\in B_n$ and put the information bits in
the channels $j\in G_n$. Here ``freezing'' means that we put a fixed
value in these positions and this value is known both to the
transmitter as well as the receiver. In fact, we are free to choose
the value and generically we will choose this value to be $0$. 
Decode the bits $U_1,\dots,U_{2^n}$ successively from $1$ to $N$.
Of course, if a bit $U_k$ is frozen, then we already know its value
and no actual decoding has to be done. Only if $U_k$ belongs to the
good set $G_n$ will we need to decode. But in this case the error
probability will by definition be very small.  The associated
computational complexity is of the order
$\mathcal{O}(n2^n)=\mathcal{O}(N\log_2N)$.\footnote{That this is
indeed the case takes some thinking.  A closer look shows that the
``graphical model'' that describes the relationship between the
input and the output, i.e., the graphical model corresponding to
the binary matrix that describes this relationship, has
$\mathcal{O}(N\log_2N)$ nodes and edges and decoding can be
accomplished by computing one message for each edge in this model.}
Since good channels are very good, the union bound on the error
probability decays, and it decays like $N2^{-\sqrt{N}}$.  Finally,
note that $\lim\limits_{n\rightarrow +\infty}|G_n|2^{-n}=1-\epsilon$.
This means that the fraction channels that we can use for information
transmission is $1-\epsilon$.  This fraction is equal to the upper
bound on the capacity we previously derived.  Hence, we have matching
upper and lower bounds on the capacity and therefore determined
capacity exactly. In addition we have found a low-complexity capacity
achieving scheme!

\subsection{Summary}
For the BEC($\epsilon$), the capacity is $1-\epsilon$. For polar
codes, the encoding and decoding complexity is of the order
$\mathcal{O}(N\log_2N)$, where $N=2^n$ is the blocklength.

We mention a final point. Assume that we want to transmit information
at a rate $R=C(1-\delta)$. Here $\delta$ represents the so-called
{\em gap} (of the rate) {\em to the capacity}. Assume further that
we want to achieve a certain fixed block probability of error. Then
what blocklength is required to achieve this goal? More precisely,
if we let $\delta$ tend to $0$, then how does the blocklength have
to scale with $\delta$?  This question was addressed by Strassen as
well as Polyanskiy, Poor and Verdu \cite{STR62,PPV10}.  The result
is that, for any code, the blocklength must grow at least as the
square of the reciprocal of the gap to capacity, i.e., $N =
\Theta(1/\delta^2)$ and there exist coding schemes which achieve
this lower bound.

How do polar codes stack up in terms of their finite-length scaling?
It was shown in \cite{HKU} that for general channels we have
\begin{equation*}
\frac{1}{\delta^{3.56}}\leq N \leq \frac{1}{\delta^7}.
\end{equation*}
For transmission over the BEC, we need $N=1/\delta^{3.67}$ and  for
the Binary Symmetric Channel (BSC) we need $N=1/\delta^{4.2}$. So
this means that polar codes require roughly the square of the
blocklength compared to optimal codes.

A BSC($\epsilon$) is a channel that takes a bit $x\in\{0,1\}$. It
flips the bit to $1-x$ w.p. $\epsilon$ and leaves it unchanged w.p.
$1-\epsilon$.  

So far we only talked about polar codes for the binary erasure
channel. But everything we mentioned can be extended to more general
channels, such as the BSC or the so-called additive white Gaussian-noise
channel (AWGNC). In the same manner as for the BEC we can construct
low-complexity capacity-achieving polar codes or such channels.

Another caveat concerning polar codes concerns the question of
``universality." Consider a polar code $\mathcal{C}_1$ for the
BEC($\epsilon$) and a polar code $\mathcal{C}_2$ for the BSC($p$).
Assume that the parameters $\epsilon$ and $p$ are chosen in such a
way that the capacities of the two channels are equal. We know from our
previous discussion that for both scenarios we can construct capacity-achieving
polar codes. This means that if we pick $N=2^n$ sufficiently large then the
fraction of good indices in both cases is close to capacity.

Denote by $B_{n,1}$ and $B_{n,2}$ the sets of good channels
corresponding to the $\mathcal{C}_1$ and $\mathcal{C}_2$, respectively.
Denote by $G_{n,1}$ and $G_{n,2}$ the sets of bad channels similarly.
It is now natural to ask if 
\begin{align*}
B_{n,1}&=B_{n,2}?, \\
G_{n,1}&=G_{n,2}?
\end{align*}
In words, we are asking if the {\em same} synthetic channels are
good for the two scenarios.  If this is the case then the code is
{\em universal}, meaning one and the same code is good for both
scenarios.

Unfortunately, the answer has been found to be negative, and so
polar codes are not universal\footnote{In fact, recent results show
that they can be made universal at the price of increasing the
blocklength.}

\section{Applications}
Before we continue and describe codes based on sparse graphs it
might be interesting to consider some standard application scenarios.
This will make it clearer what range of parameters is typically of
interest.

Consider transmission over the AGWNC. That is,
\begin{align*}
y_i=h_ix_i+z_i,
\end{align*}
where $x_i\in\{-1,+1\}$ is the bit that we want to transmit, $y_i$
is the received value, $z_i\sim\mathcal{N}(0,\sigma^2)$ (Gaussian
noise of zero mean and unit variance), and $h_i$ is the so-called
fading coefficient, describing the path loss of signal strength
caused by the transmission medium.

In {\em wireless} transmission settings, we are using electromagnetic
waves emitted and captured by {\em antennas}, as the transmission
medium.  In practice, the following values are typical
\begin{itemize}
\item blocklength: $N\sim 10^3-10^4$ bits,
\item rate: $R\sim 0.5$,
\item block error probability: $P_B\sim 10^{-2}$,
\item throughput: $10^4-10^6$ bits/sec,
\item processing power consumption: $10$mW.
\end{itemize}
Thus, we have at our disposal about $10^{-7}$ Joules to process one bit. 

Another transmission scheme is that over the BSC. In that case,
\begin{align*}
y_i=x_i \oplus z_i,
\end{align*}
where $x_i\in\{0,1\}$, $z_i\in\{0,1\}$, and P($z_i=1$)$=p$.

This is a first-order approximation to model optical transmission. In such settings, the following values are typical
\begin{itemize}
\item blocklength: $N\sim 10^4-10^6$ bits,
\item rate: $R\sim \frac{239}{255}$ for historical reasons,
\item bit error probability: $P_B\sim 10^{-15}$, basically ``one error per day",
\item throughput: $100$ Gbits/sec 
\item processing power consumption: $100$W.
\item interchip data rate: $5\times 10^{12}$ bits/sec (within chip, for message passing),
\end{itemize}
Thus, we have at out disposal about $10^{-9}$ Joules to process one bit, quite a limited quantity. 

\subsection{Metrics}
How can one ``measure" codes so that we can compare various competing schemes in a meaningful manner? 
The following are some useful metrics.
\begin{itemize}
\item {\bf Construction complexity}: How difficult is it to find a code? For polar codes this can be done quite efficiently. 
\item {\bf Encoding and decoding complexity}: How many operations
do we need to encode and decode one information bit? As we have
seen, for polar codes both encoding and decoding can be done in
$\mathcal{O}(N \log_2N)$ (real) operations, where $N$ is the
blocklength. This is also efficient.
\item {\bf Finite length performance}: What blocklengths do we need
in order to get ``close'' to capacity. This is one of the few
weaknesses of polar codes. We have seen that the required blocklength
is roughly the square of what is optimally achievable.
\item {\bf Throughput}: How many bits can we decode per clock cycle.
For some high speed applications such as for optical transmission
on the backbone network this is important. The standard decoder for
polar codes is inherently sequential and so does not have a very
high throughput. But it can made more parallel if we are willing
to pay a higher processing cost.  \item {\bf Universality}: Is one
and the same code good for many channels? Standard polar codes are
not universal, but they can be made universal if we are willing to
consider longer codes.
\item {\bf Proofs}: How simple is it to explain the scheme? Polar
codes are by far the simplest of all known capacity-achieving
schemes. In addition they have an explicit construction rather than
only probabilistic guarantees.  \end{itemize}

\section{Low-Density Parity-Check Codes} 
\subsection{Linear Codes}
Linear codes are codes so that the (weighted) sum of any two codewords
is again a codeword. As a consequence, such codes have a compact
algebraic description, either as the image of a linear map or as
the kernel of a linear map.  In the first case we typically consider
the so-called {\em generator} matrix $G$ and we represent the code
as the space spanned by the rows of $G$. More precisely, let $G$
be an $k \times n$ matrix over a field $\mathbb{F}$. Here, $n$ is
the {\em blocklength}\footnote{In the previous chapter the blocklength
was denoted by $N$ and $N=2^n$, where $n$ denoted the number of
polar steps. For the most part we will revert now to the more
standard notation where $n$ denotes the blocklength except when we
talk about polar codes. We hope that the resulting confusion will
stay bounded.} and $0 \leq k \leq n$ is the {\em dimension} of the
code.  Although more general cases are possible and indeed can be
useful, we will restrict our discussion to the case where $\mathbb{F}$
is the binary field. The code {\em generated} by $G$ is then 
\begin{equation}
C(G) = \{x \in \mathbb{F}^n : x = uG, u \in \mathbb{F}^k\} = \{x \in \mathbb{F}^n : Hx^T=0\}.
\end{equation}
Here the second representation of the code is in terms of the kernel
of the so-called {\em parity-check} matrix $H$.  Note that if $G$
has rank $k$ (and thus $|C(G)| = 2^k$), then by the rank-nullity
theorem $H$ has rank $n-k$.

As an important example, {\em polar} codes that we discussed in the
prequel, are linear codes.  We can find the generator matrix
corresponding to polar codes by starting with the binary matrix
\begin{align*}
G_1=\left[
\begin{array}{cc}
1 & 0 \\
1 & 1
\end{array}
\right]
\end{align*}
Let $G_n$, $n \in \mathbb{N}$, be the $n$-th Kronecker product of $G_1$. The generator matrix corresponding
to a polar code of length $N=2^n$ is then the matrix which corresponds to picking those rows of $G_n$ which
correspond to the ``good'' channels.
Almost all codes used in practice are linear. This has two reasons.
First, it can be shown that for most scenarios linear codes suffice
if we want to get close to capacity. Secondly, linear codes are
typically much easier to deal with in terms of complexity.
\index{MAP}
\subsection{MAP decoding}
In order to decode the output of the noisy channel, an appropriate choice is the Maximum-A-Posteriori (MAP) estimator:
\begin{equation}
\label{MAP}
\hat x^{MAP} =\underset{c\in C(G)}{\text{argmax}}P_{X|Y}(x|y) = \underset{c\in C(G)}{\text{argmax}} P_{Y|X}(y|x) \frac{P_X(x)}{P_Y(y)} = \underset{c\in C(G)}{\text{argmax}} P_{Y|X}(y|x) P_X(x)
\end{equation}
where $P_X(x)$ is the prior over X and $P_{Y|X}(y|x)$ is the
likelihood of the output $Y$ of the noisy channel given $X$. The
MAP decoder outputs the mode of the posterior distribution and thus
minimizes the block-error probability. This is why \index{Y} we would like
to implement it. (In order to achieve capacity it is in fact not
necessary to do MAP decoding).

\subsection{Low Density Parity Check codes}
Low-density parity-check codes (LDPC) are linear codes defined by
a parity-check matrix $H$ that has few non-zero entries, more
precisely, the number of non-zero entries only grows linearly in
the dimension $n$ of the matrix.

\begin{figure}
\centering
\begin{minipage}{.5\textwidth}
\centering
\includegraphics[width=0.6\linewidth]{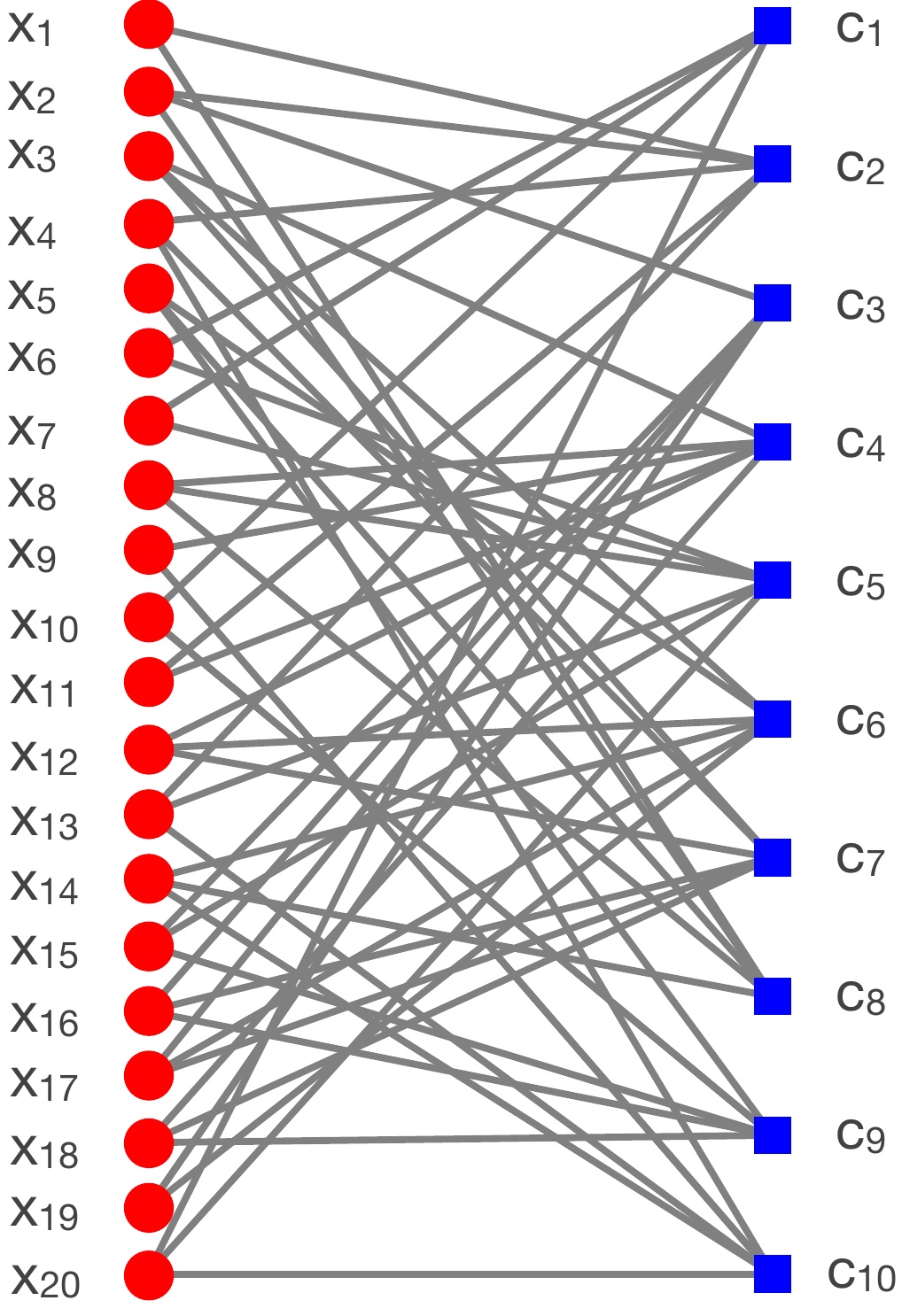}
\captionof{figure}{Factor graph of a LDPC with N=20 variables and 10 factors.}
\label{fig1Jean}
\end{minipage}
\begin{minipage}{.47\textwidth}
\centering
\includegraphics[width=0.82\linewidth]{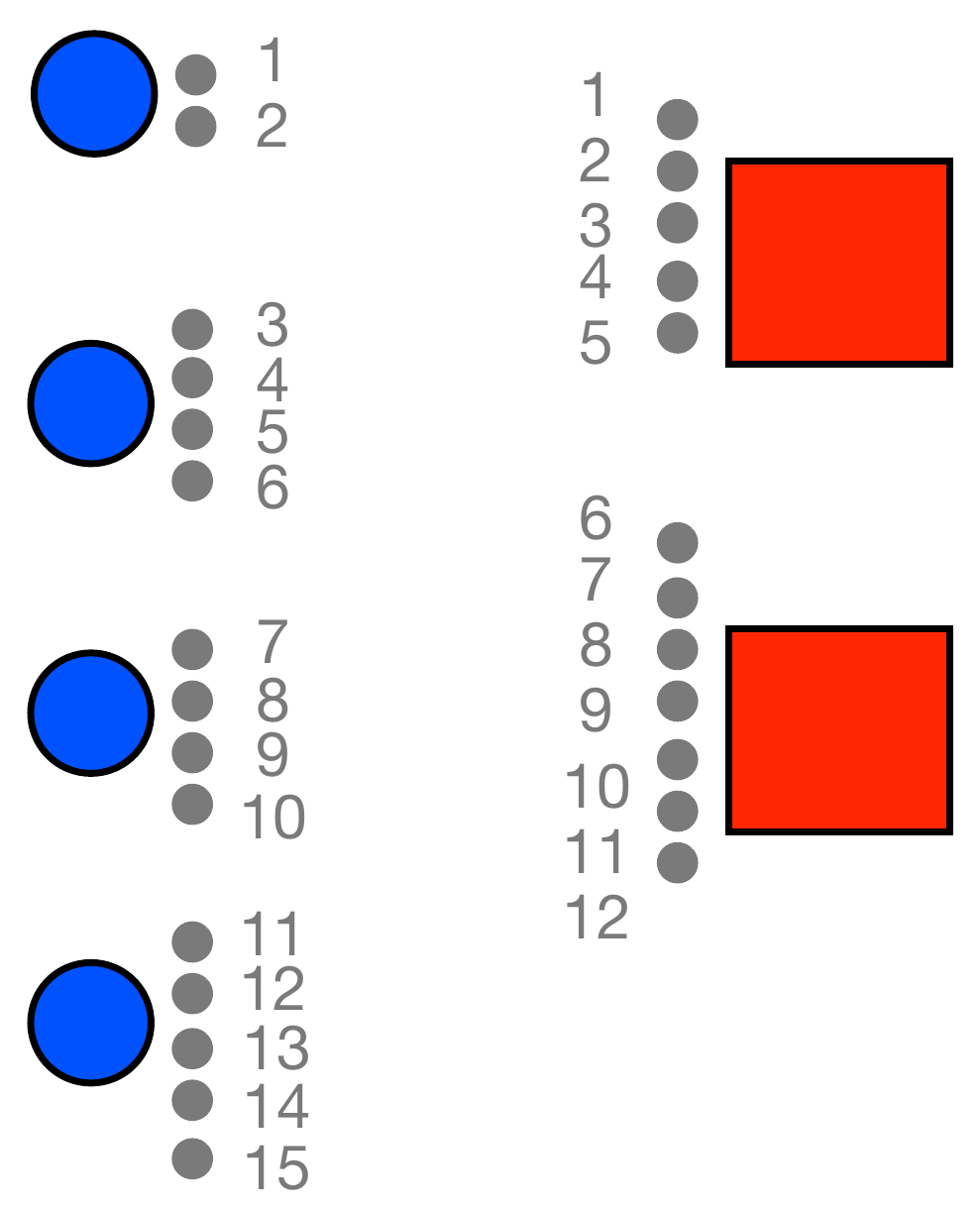}
\captionof{figure}{Instance of a (3,6)-random factor graph, where 3 (resp., 6) is the average degree of nodes (resp., factors).}
\label{fig2Jean}
\end{minipage}
\end{figure}

A particularly useful description is in term of a factor graph (see
Fig.~\ref{fig1Jean}).  In this factor graph there are $n$ {\em
variable} nodes representing the components of the codeword and
there are $n-k$ factor nodes, each representing one of the $n-k$
linear constraints implied by the parity-check matrix. There is an
{\em edge} between a factor node and a variable node if that
particular node participates in the constraints represented by the
factor node. Since the parity matrix $H$ is {\em sparse}, the number
of edges in this factor graph only grows linearly in the length of
the code.

The code is then the set of all binary $n$-tuples that fulfill each
of the $n-k$ constraints.  As an example, looking at Fig.~\ref{fig1Jean} the first constraint is $C1 = \mathbb{I}(x_6 \oplus
x_7 \oplus x_{10} \oplus x_{20} = 0)$ where $\mathbb{I}(\bullet)$
is the indicator function.
The code rate, i.e., the  fraction of information bits contained
in the $n$ transmitted bits, is equal to $R\ge R_{\text{design}} =
\frac{\#\text{variables} - \#\text{factors}}{\#\text{variables}}$.
In general, $R_{\text{design}}$ is only a lower bound on the actual
rate since some of the constraints can be linearly dependent.  But
it was shown by \cite{RateRegularLDPC}, that with
high probability the rate of a randomly chosen regular code is very
close to this lower bound (see also \cite{MCT}). A regular code one where
all variables have constant degree $d_l$ and all check
nodes have degree $d_r$.

\subsection{Configuration model}
One possible way of generating a $(d_l, d_r)$-regular code is to
define an {\em ensemble} of such codes and to create a specific
instance by sampling uniformly at random from this ensemble. A
canonical way of achieving this is via the so-called {\em configuration}
model. In this model, associate $d_l$ ``sockets'' to each variable
node and $d_r$ sockets to each check node. Note that there are in
total $n d_l$ variable node sockets and an equal number, namely
$(n-k) d_r$ check-node sockets. We get from this model a graph by
picking uniformly at random a permutation on $n d_l$ elements and
by matching up sockets according to this permutation.  This ensemble
is convenient for practical implementations, since it is easy to
sample from, and it is also well-suited for theoretical analysis.

For applications it  is also important to be able to have nodes of various degrees.
The corresponding ensembles of codes are called {\em irregular} ensembles. 
To specify such an ensembles we need to specify how many nodes there are of what degree,
or equivalently, how many edges there are that connect to nodes of various degrees.

One useful representation of the statistics of the degrees is in
terms of a polynomial representation. For example, the polynomials
corresponding to Fig.~\ref{fig2Jean} are:
\begin{eqnarray}
\Lambda(x) &=& \frac{1}{4}x^2 + \frac{2}{4}x^4 + \frac{1}{4}x^5, \ \lambda(x) = C \frac{d\Lambda(x)}{dx} = \frac{2}{15}x + \frac{8}{15}x^3 + \frac{5}{15}x^4 \nonumber\\
P(x) &=& \frac{1}{2}x^5 + \frac{1}{2}x^7, \ \rho(x) = C \frac{dP(x)}{dx} \frac{5}{12}x^4 + \frac{7}{12}x^6
\end{eqnarray}
Here, $\Lambda$ ($P$) is the normalized distribution from the node
perspective (where $\Lambda$ specifies the variable node degrees
and $P$ specifies the check node degrees).  The coefficient in front
of $x^i$ is the fraction of nodes of degree $i$.  The normalized
derivatives of these quantities, namely $\lambda$ ($\rho$), represent
the same quantities but this time from the perspective of the edges,
i.e., the represent the probabilities that a randomly chosen edge
is connected to a node of a particular degree.

\index{MAP}
\subsection{From Bit MAP to Belief propagation decoding}
\begin{figure}
\centering
\includegraphics[width=5.5cm]{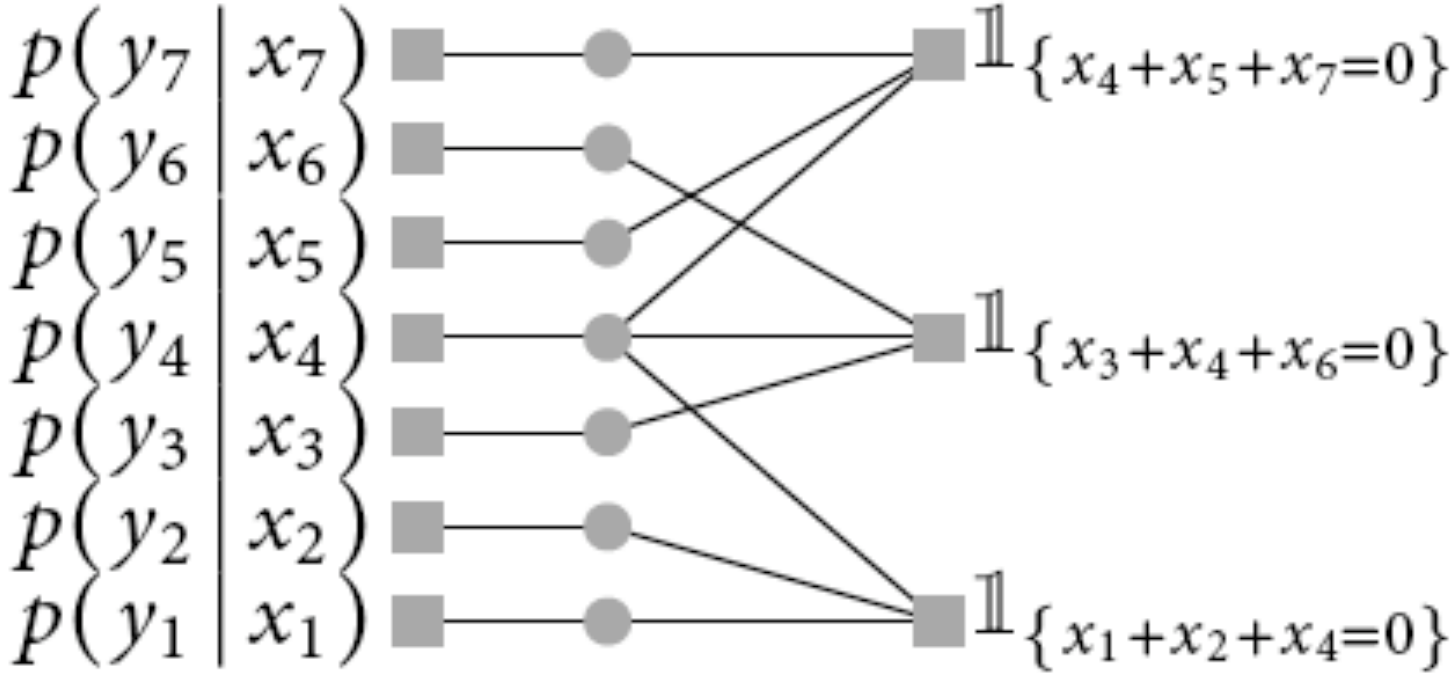} \caption{Instance of a random factor graph. The BP decoder allows to estimate the marginal or mode of each input components.\label{fig3Jean}}
\end{figure}

\index{Belief propagation}
For the BEC, bit MAP decoding can be done by solving a system of
linear equations, i.e, in complexity $O(n^3)$: one must solve $Hx
= 0$, i.e. $H_\epsilon x_\epsilon \oplus
H_{\bar\epsilon}x_{\bar\epsilon}=0$, where $H_\epsilon$ is submatrix
of the parity-check matrix spanned by the columns corresponding to
the erased components of $x$, i.e, $x_\epsilon$, and $H_{\bar\epsilon}$
is the complement. Thus $H_\epsilon x_\epsilon =
H_{\bar\epsilon}x_{\bar\epsilon}=s$ has to be solved to find back
the missing part $x_\epsilon$.  But we are interested in an algorithm
that is applicable for general binary-input memoryless output-symmetric
(BMS) channels, where MAP decoding is typically intractable. We
therefore consider a message-passing algorithm which is
applicable also in the general case. More precisely, we consider
the sum-product (also called Belief-Propagation (BP)) algorithm.
\index{Belief propagation}
This algorithm performs bit MAP decoding on codes whose factor graph
is a tree, and performs well on locally tree like graphs such as
the random ones (see Fig.~\ref{fig3Jean})
From \eqref{MAP} we get
\begin{figure}
\centering
\includegraphics[width=5cm]{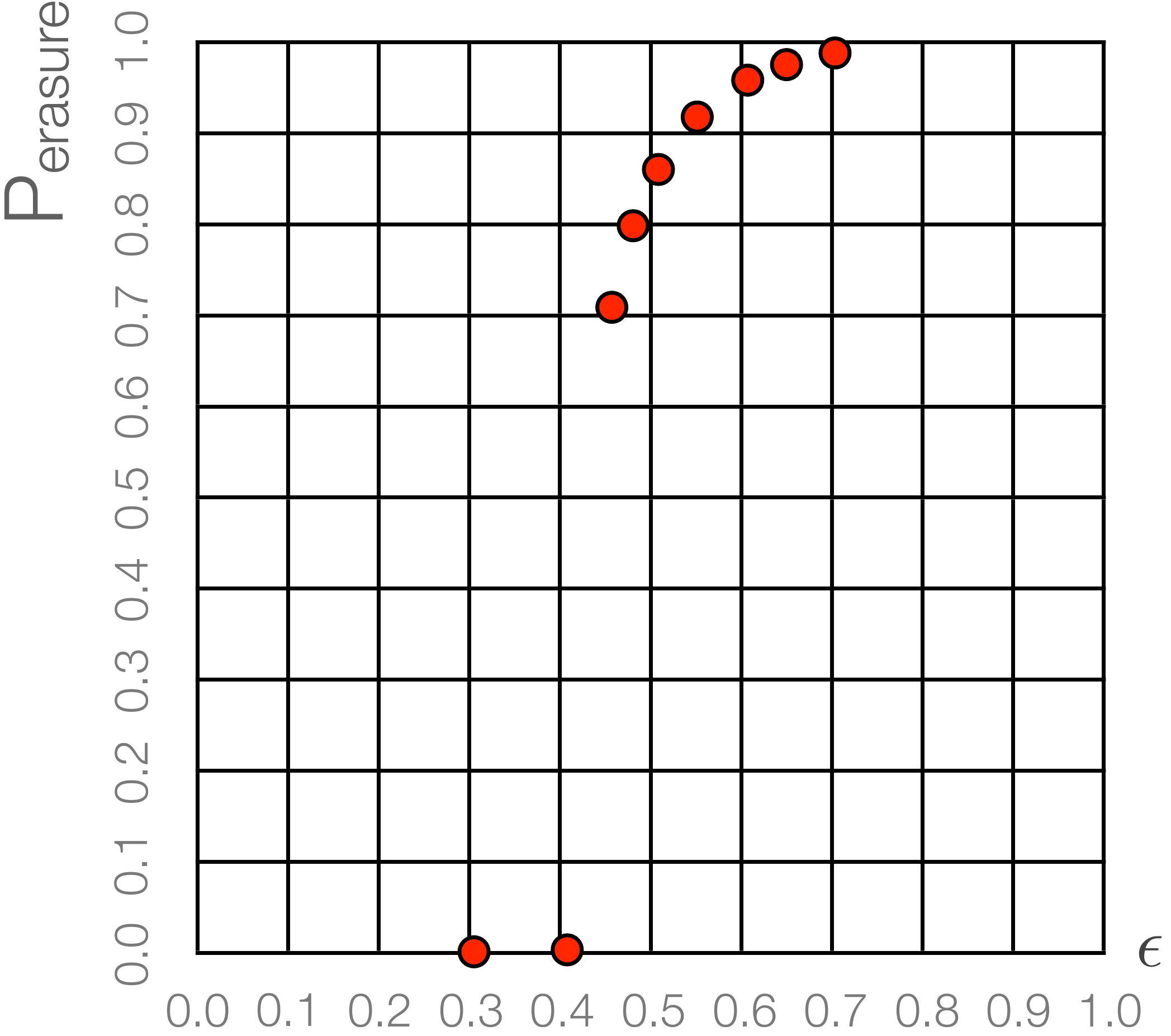} \caption{Performance of BP decoding over a BEC channel.\label{fig33Jean}}
\end{figure}
\begin{equation}
\hat x_i^{\text{\small MAP}} = 
\underset{x_i\in\pm 1}{\text{argmax}} \sum_{\{x_j:j\neq i\}} \left(\prod_j p(y_j|x_j)\right) \mathbb{I}\left(x\in \mathcal{C}\right)
\end{equation}
BP is a message passing algorithm that finds a fixed point to the
following set of equations (here for the case of a parity check
code):
\begin{eqnarray}
m_{i\to\mu}(x_i) &=& \frac{p(y_i|x_i)}{z_{i\to\mu}} \prod_{\nu\in \partial_i\backslash \mu } \hat m_{\nu\to i}(x_i) \\
\hat m_{\mu\to i}(x_i) &=& \frac{1}{\hat z_{\mu\to i}} \sum_{\{x_j:j\in \partial_\mu\backslash i \}}\mathbb{I}\left(\left[\bigoplus_{\{x_j: j \in \partial_\mu\backslash i\}} x_j \right]\oplus x_i = 0\right) \prod_{\{x_j : j \in \partial_\mu\backslash i\}} m_{j\to\mu}(x_j)\nonumber
\end{eqnarray}
 where $\partial_\mu\backslash i$ stands for the ensemble of variable
 indices of the variables that are neighbors of factor $\mu$ except
 $i$, the messages $\{m_{i\to\mu},\hat m_{\mu\to i}\}$ are the
 so-called cavity messages (which are probability distributions,
 $\{z_{i\to\mu},\hat z_{\mu\to i}\}$ are the normalization constants)
 from which we can infer their most probable state by maximization
 of the marginals $\{m(x_i)\}$ allowing bit MAP decoding:
\begin{equation}
m(x_i) =  \frac{1}{z_i}\prod_{\nu\in \partial_i} \hat m_{\nu\to i}(x_i)
\end{equation}
What is the performances of the BP algorithm on the BEC? Here is the experiment
we consider. Fix the ensemble. In the above example it is the (3,
6)-regular ensemble. Now pick very long instances of this ensemble.
Pick a random codeword and transmit over a BEC with erasure probability
$\epsilon$. Run the BP decoder until convergence. Record the error
probability and average over many instances. Plot the average
bit-error probability versus $\epsilon$. Naturally, as $\epsilon$
decreases the error probability decreases. What is most interesting
is that at some specific point we see a jump of the error probability
from a non-zero value down to zero. This is the {\em BP threshold} (see
Fig.~\ref{fig33Jean}).
 \subsection{Asymptotic Analysis: Density Evolution (DE)} 
 \begin{figure}
 \centering
 \includegraphics[width=12cm]{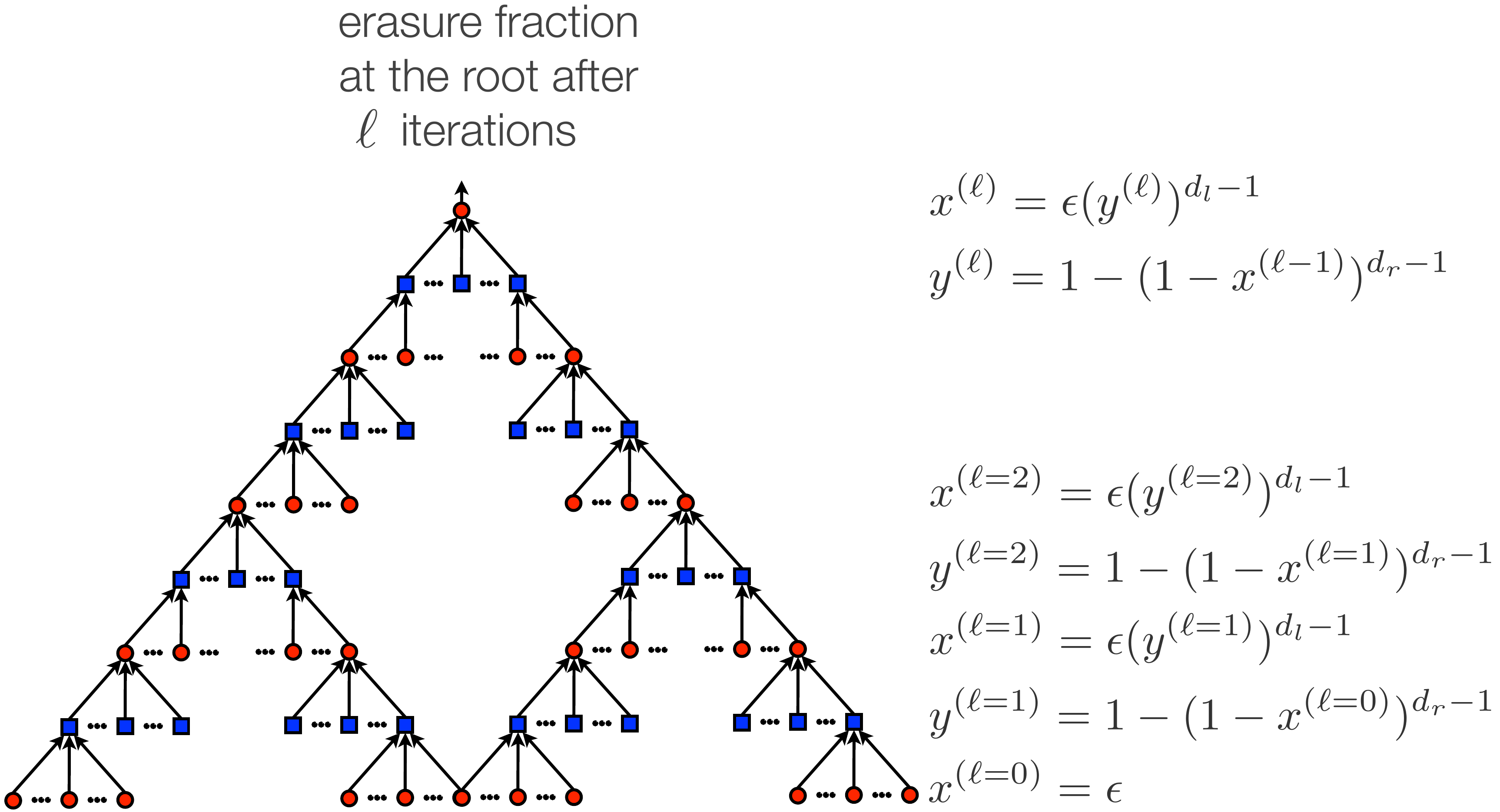} \caption{Representation of the DE dynamics over an infinite tree, allowing the computation of the asymptotic probability of decoding at the root.\label{fig4Jean}}
 \end{figure}
\begin{figure}
\centering
\includegraphics[width=7cm]{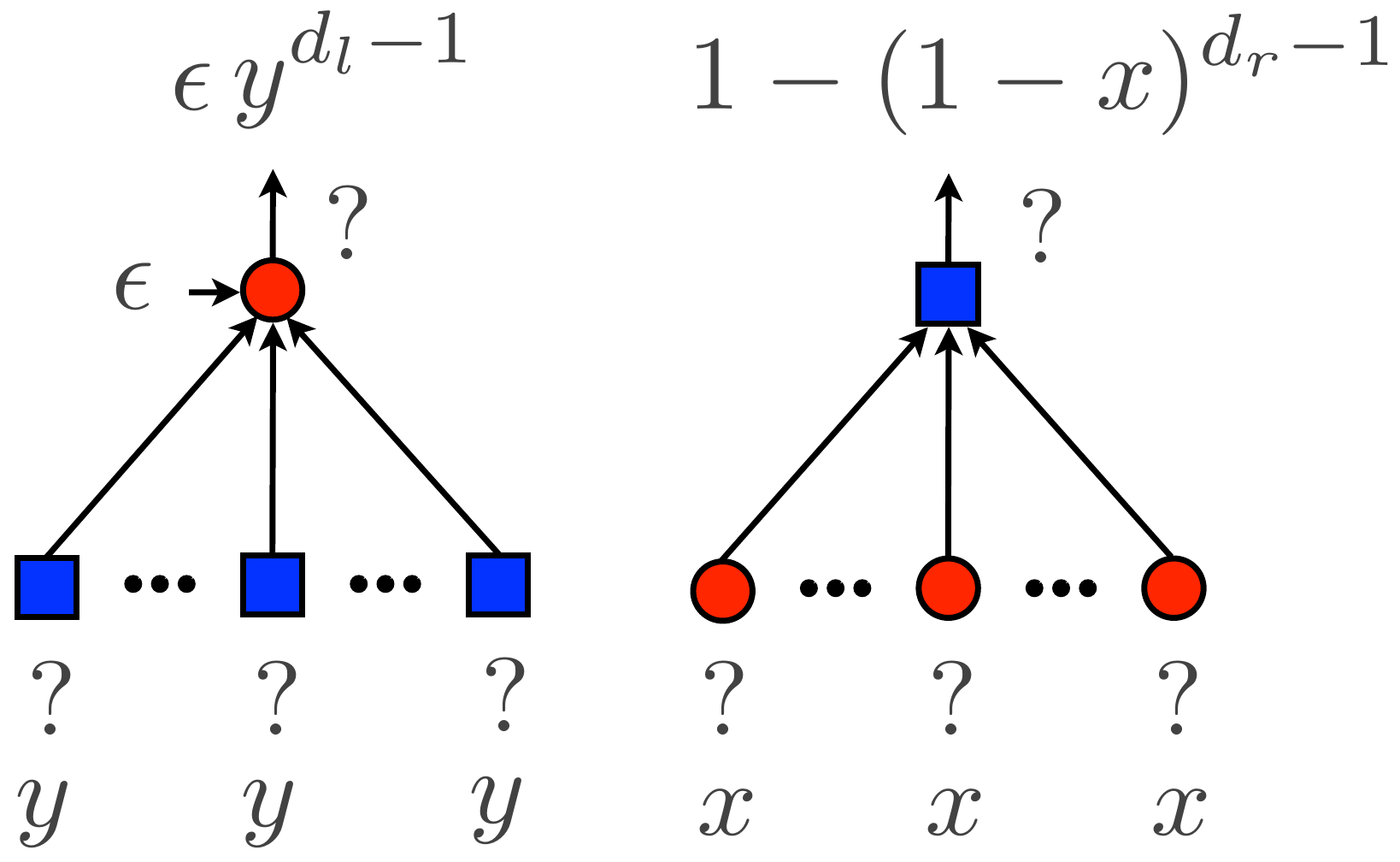} \caption{DE iteration over a factor and a node. \label{fig5Jean}}
\end{figure}
\begin{figure}
\centering
\begin{minipage}{.5\textwidth}
\centering
\includegraphics[width=0.6\linewidth]{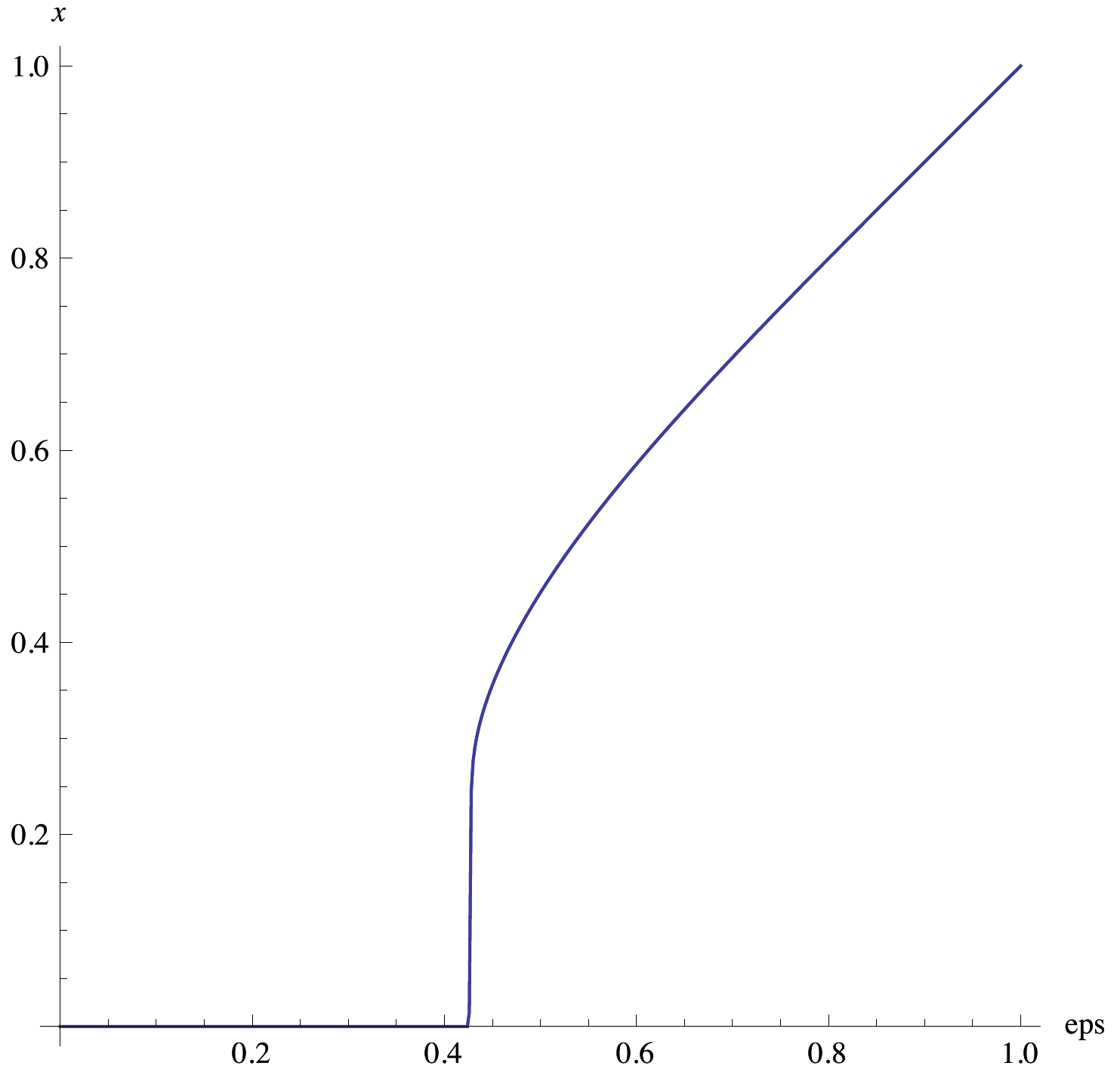}
\captionof{figure}{DE prediction for the fraction of lost bits after BP decoding for the BEC as a function of the erasure probability.}
\label{fig7Jean}
\end{minipage}
\begin{minipage}{.47\textwidth}
\centering
\includegraphics[width=1\linewidth]{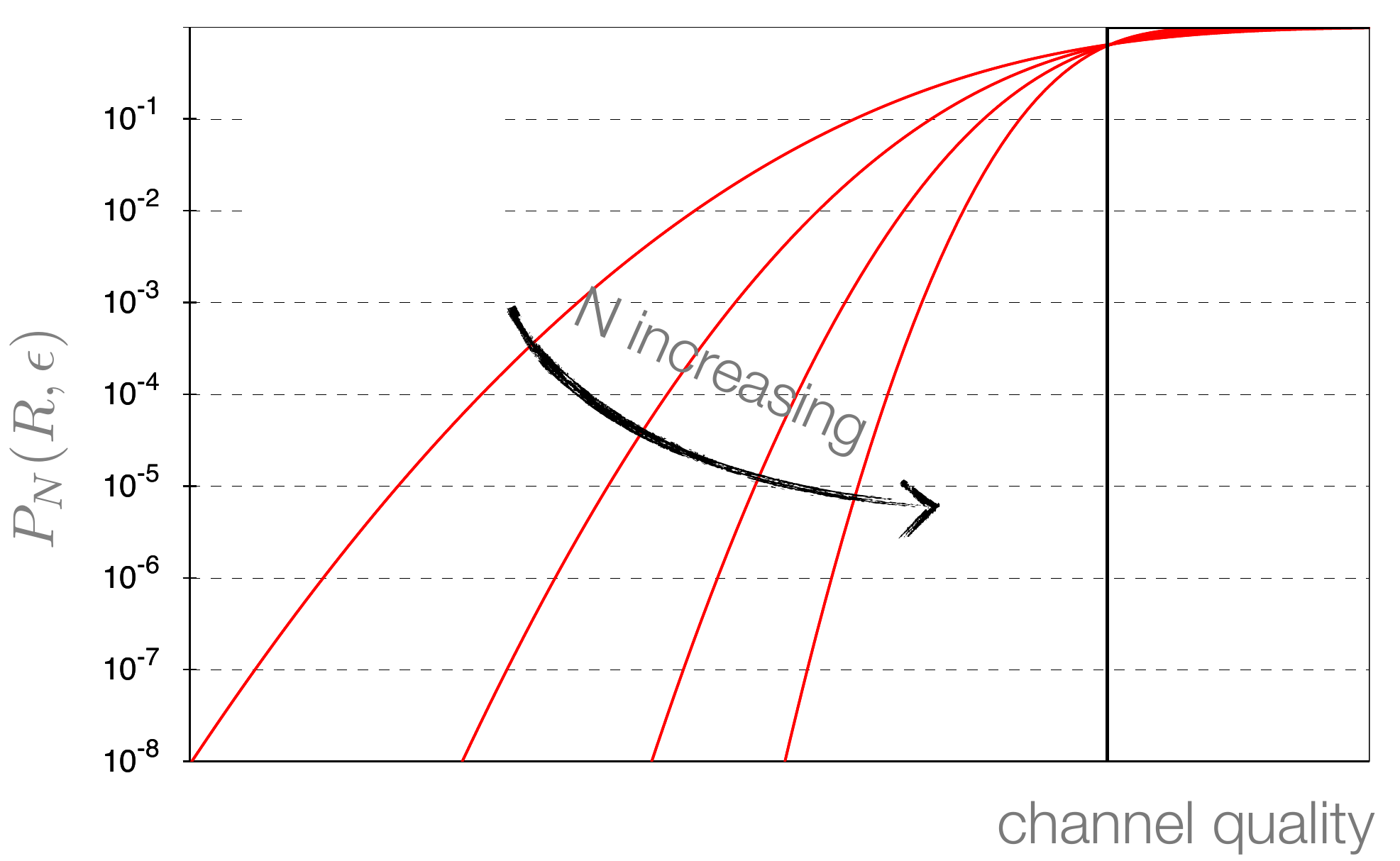}
\captionof{figure}{The experimental curve getting closer to the BP threshold as N increases.}
\label{fig11Jean}
\end{minipage}
\end{figure}

Density evolution is a general method that allows us to analyze
decoding in the limit where the number of nodes and factors both
become large but their ratio remains constant. We do this by looking
at how the erasure probability behaves at each of the two types of
the nodes. Consider a $(d_l, d_r)$-regular code, i.e., every variable
node has degree $d_l$ and and every check node has degree $d_r$.
We focus on the BEC. At the variable node, if there is an incoming
message which is not an erasure, then the variable node is exactly
determined. This is because we are transmitting over the BEC and
either we have perfect information or we have absolutely useless
information. On the check node side, even if only one incoming
message is an erasure, the check node output has no way knowing
whether it is 0 or 1. Denoting $y$ ($x$) the probabilities that a
factor (node) is undetermined, we obtain Fig.~\ref{fig5Jean}
giving the probabilities for a node (factor) to output no information
after one iteration.\\ So if we perform $l$ iterations we get a
sequence of erasure probabilities as shown in Fig.~\ref{fig4Jean}.
This is how Gallager analyzed LDPC codes. Luby and. al. used a
somewhat different procedure. In their analysis they look at the
so-called peeling decoder. This decoder is entirely equivalent to
the BP decoder (when transmitting over the BEC). In this decoder,
as long as there is a degree-one check node, we use this check node
to determine one more bit and then remove the used check node as
well as the determined variable. We then follow the evolution of
the graph. This can be done by writing down a system of differential
equations. This method is called the Wormald method \cite{WormaldMethod}.\\
Note that in the density evolution approach we assume that we first
fix the number of iterations and let the length of the code tend
to infinity (so that there are no loops in the graph up to the
desired size). We then let the number of iterations tend to infinity.
In the Wormald approach on the other hand we take exactly the
opposite limit. Luckily both approaches give exactly the same
threshold: DE corresponds to the limit $\lim_{l\to\infty}\lim_{n\to\infty}$
but in fact we can take the limit in any order, or jointly, and
we'll always get the same threshold: the approach is robust. The
density evolution is decreasing and bounded from below and will
thus converge. For large codes, the behavior of almost all of them
in the ensemble is accurately predicted by DE: it is the concentration
property. DE can be applied to the BEC to predict the fraction of
bits that cannot be recovered by BP decoding as a function of the
erasure probability (see Fig.~\ref{fig7Jean}). It is predicted
that there exist a critical threshold ($\epsilon \approx 0.429$ for
the BEC) under which BP will recover the full codeword and above
which it becomes impossible to recover everything. It perfectly
matches the experimental threshold (Fig.~\ref{fig33Jean}) but the
curves are different. We will understand why \index{Y} in the next section.

\subsection{EXIT curves}
Instead of plotting the $x$-value on the vertical axis it is often
more convenient to plot the EXIT value, see Fig.~\ref{fig8Jean}.
The EXIT value has a simple interpretation. It is the error probability
of the best estimate we can do using all the internal messages at
a node but without the channel observation at this bit. This is why \index{Y}
we have $y$ to the power $d_l$ and not $d_l-1$ but we do not have
the factor $\epsilon$ corresponding to the channel erasure fraction.
We will see soon why \index{Y} the EXIT value is the right quantity to plot.
Rather than running the recursion we can right away find the value
to which the recursion converges. This is because this final value
must be a solution to the fixed-point (FP) equation $x=f(\epsilon,
x)$, where $f(\cdot)$ denotes a recursive DE equation.
The forward fixed points of DE (see Fig.~\ref{fig9Jean}), which follows the true decoding dynamics and with initial condition $x^{(l=0)} = \epsilon$ are:
\begin{eqnarray}
y^{(l)} &=& 1-(1-x^{(l-1)})^{d_r-1}\\
x^{(l)} &=& \epsilon(y^{(l)})^{d_l - 1}\\
x^{(l)} &=& \epsilon(1-(1-x^{(l-1)})^{d_r-1})^{d_l-1}
\end{eqnarray}
Then, the fixed points of DE (see Fig.~\ref{fig10Jean}) are obtained by removing the time step index:
\begin{eqnarray}
x&=&\epsilon(1-(1-x)^{d_r-1})^{d_l-1} \nonumber \\
\epsilon &=& \frac{x}{(1-(1-x)^{d_r-1})^{d_l-1}} \label{equ:fixedpoint}
\end{eqnarray}
Note that there are in general several values of $x$ which satisfy
the FP equation for a given $\epsilon$, but there is always just a
single value of $\epsilon$ for a given $x$, which is easily seen
by solving for $\epsilon$ from the FP equation above. This makes
it easy to plot this curve. But note also that in this picture we
have additional fixed points. These fixed points are unstable and
we cannot get them by running DE. The previous DE equations can be
easily extended to the irregular graph case:
\nopagebreak 
\begin{eqnarray*}
x^{(l=0)} &=& \epsilon \\
y^{(l)} &=& 1-\rho(1-x^{(l-1)})\\
x^{(l)} &=& \epsilon\lambda(y^{(l)})\\
x^{(l)} &=& \epsilon\lambda(1-\rho(1-x^{(l-1)}))\\
\end{eqnarray*}
These distributions $\rho()$ and $\lambda()$ can be optimized over by finite size scaling techniques, in order to reach capacity of the channel in the large blocklength limit. For example we can take a family of the form:
\begin{equation*}
\lambda_\alpha(x) = 1-(1-x)^\alpha, \ \rho_\alpha(x) = x^{\frac{1}{\alpha}}
\end{equation*}
and try to find the best parameter $\alpha$ such that the error probability decreases as fast as possible to zero below the channel capacity as N increases. In addition, the distributions must verify the matching condition:
\begin{figure}
\centering
\begin{minipage}{.5\textwidth}
\centering
\includegraphics[width=1\linewidth]{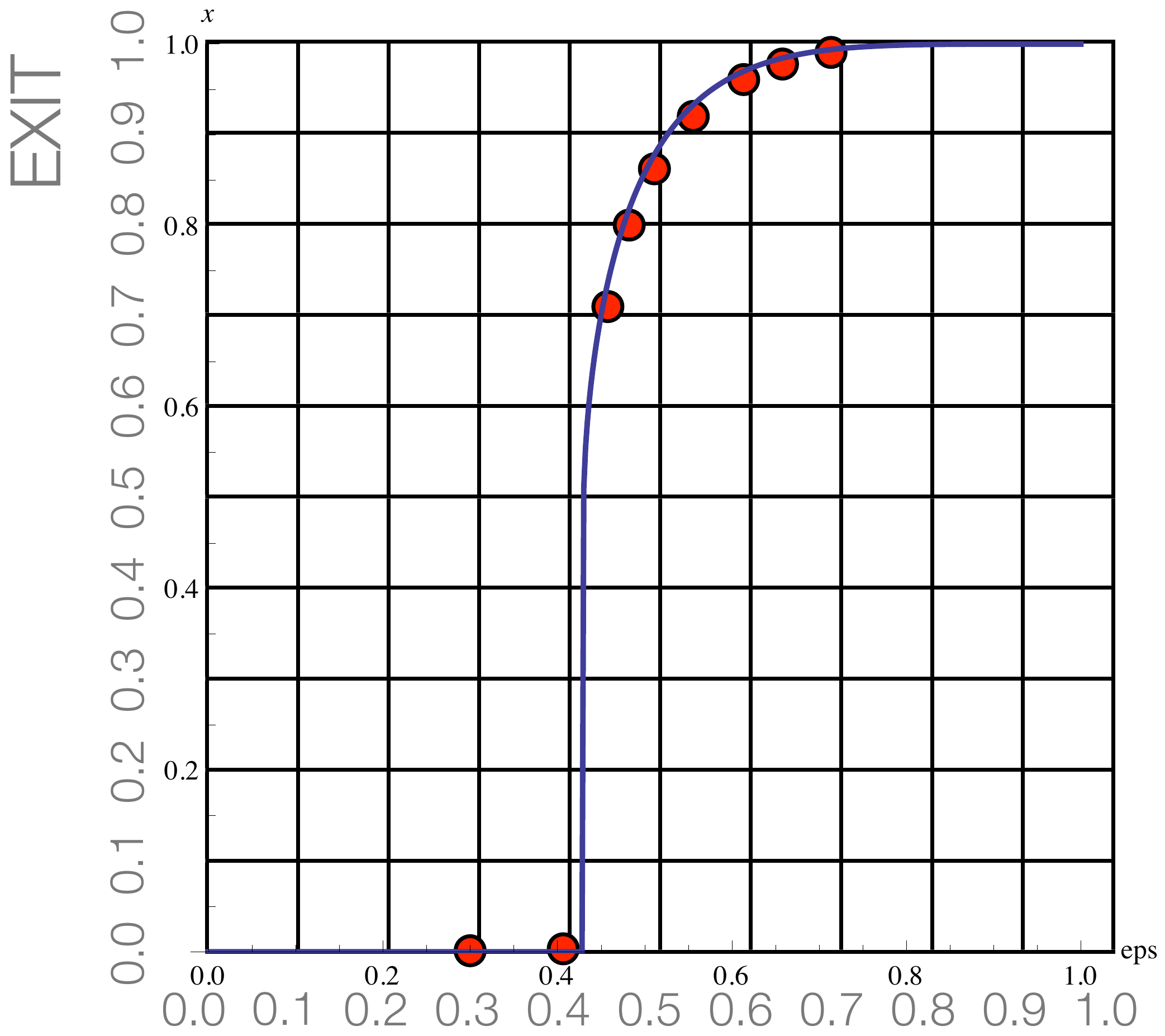}
\captionof{figure}{DE forward (stable) fixed points.}
\label{fig9Jean}
\end{minipage}%
\begin{minipage}{.47\textwidth}
\centering
\includegraphics[width=1\linewidth]{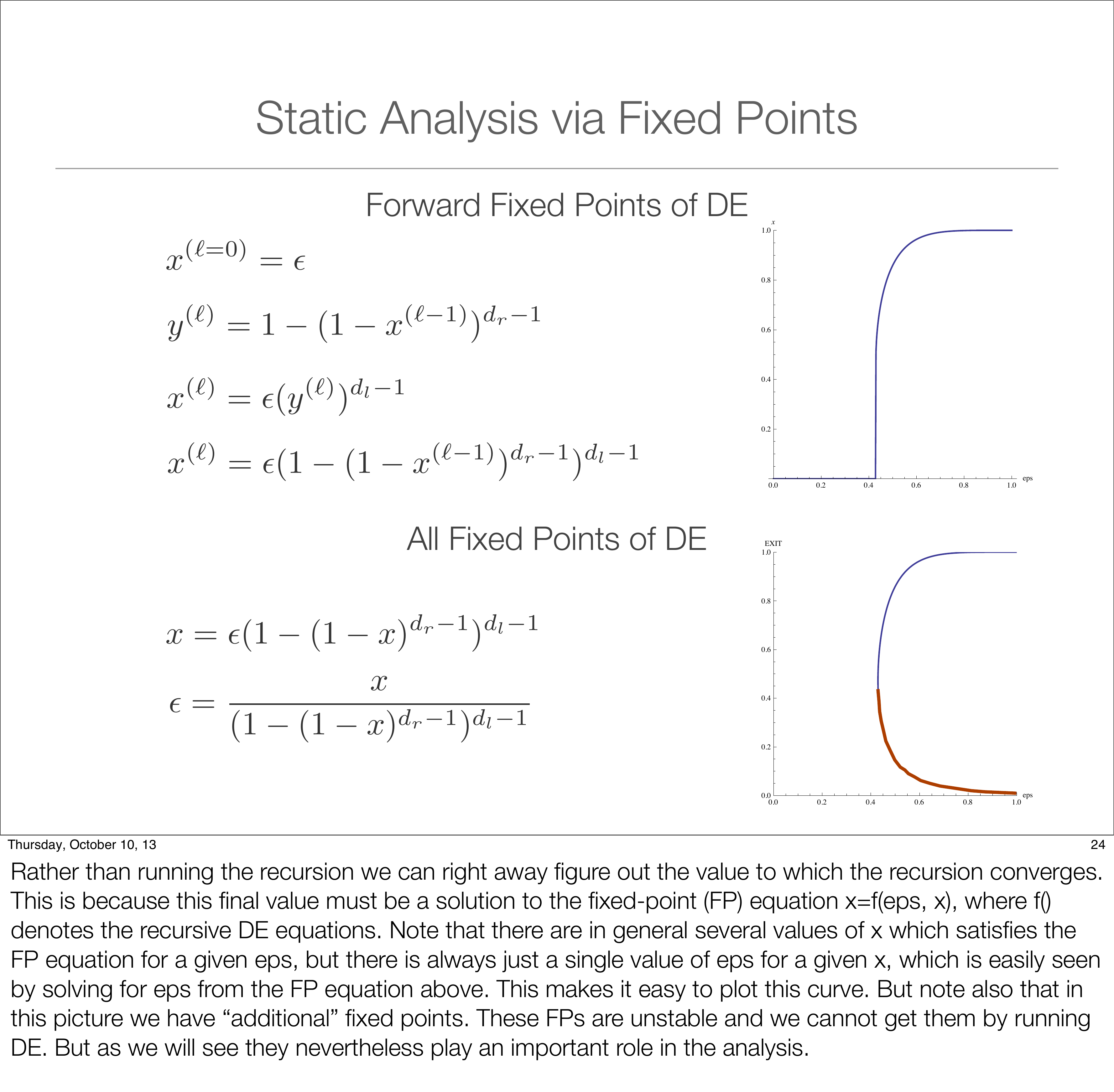}
\captionof{figure}{All DE fixed points.}
\label{fig10Jean}
\end{minipage}
\end{figure}

\begin{figure}
\centering
\includegraphics[width=7cm]{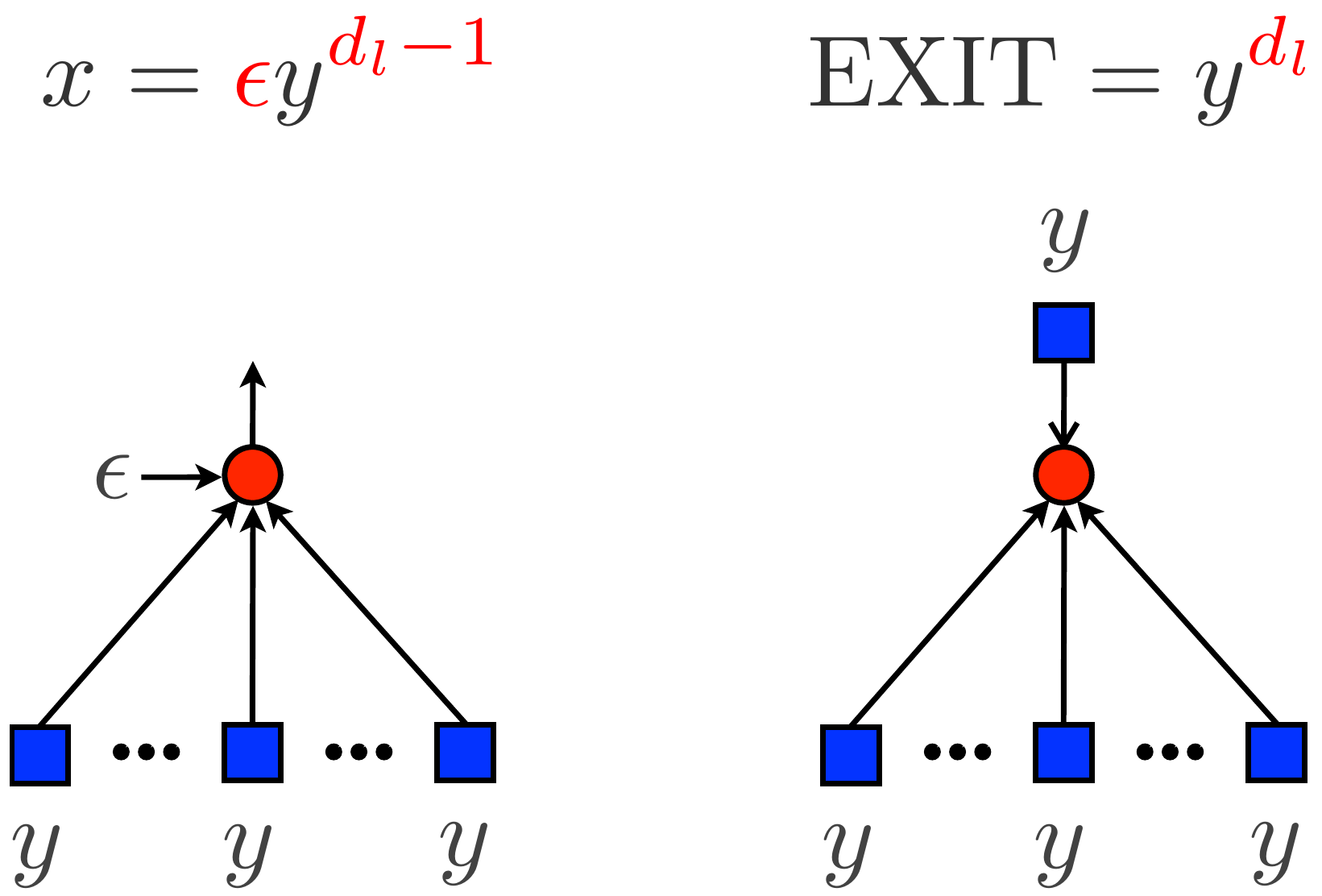} \caption{The x-value versus the EXIT value.\label{fig8Jean}}
\end{figure}
\begin{equation*}
\epsilon \lambda(1-\rho(1-x))-x\le0
\end{equation*}
Capacity achieving degree distributions should verify the strict conditions: $\epsilon \lambda(1-\rho(1-x))-x=0$ and have an average degree $\to \infty$. For instance, we can write $\lambda(x)=\sum_i w_i x^{i-1}$ with $\sum_i w_i=1$, $w_i\ge0$. In this case $\lambda()$ can be inverted and the matching condition becomes:
\begin{eqnarray*}
\int_0^\epsilon 1-\rho(1-x)dx &\le& \int_0^\epsilon \lambda^{-1}(\frac{x}{\epsilon})dx\\
\epsilon - \frac{1}{O_R}+\int_0^{1-\epsilon}\rho(x)dx&\le& \epsilon(1-\frac{1}{O_L})\\
\end{eqnarray*}
where $O_L=\int_0^\epsilon \lambda(x)dx$ is the average node degree and $O_R=\int_0^\epsilon \rho(x)dx$ is the average check degree.
\begin{equation}
\to \epsilon \le \frac{O_L}{O_R}\left(1-\frac{\int_0^{1-\epsilon}\rho(x)dx}{\int_0^{1}\rho(x)dx}\right)
\end{equation}
Again, the capacity is reached only in the strict equality case. In the case where there are n nodes and m checks in the graph, the condition $O_Ln=O_Rm$ must be true, then the rate $R = \frac{n-m}{m}=1-\frac{m}{n}=1-\frac{O_L}{O_R}=1-\epsilon_{Sh}$ where $\epsilon_{Sh}$ is the Shannon threshold satisfying $\epsilon_{Sh}=\frac{O_L}{O_R}$. It implies for the matching condition:
\index{Shannon}
\begin{equation}
\epsilon\le \epsilon_{Sh}\left(1-P(1-\epsilon_{Sh})\right)
\end{equation}
where $P$ is a polynomial that approaches 0 as $O_R\to\infty$.
\subsection{Some basic facts}
What we saw does not only work for the BEC but for a large class
of practically relevant channels. Only for the BEC we do have a
proof that these codes achieve capacity. For the general case we
need to optimize numerically. So far we looked at ensembles and
excluded many practical concerns. To find a particular code for a
standard much care and work is needed. These are the codes which
are these days included in standards. Codes are not universal but
need to be constructed with a particular channel in mind.

\section{Spatially Coupled Codes}
\index{Spatial coupling}
So far we have discussed the simplest form of LDPC ensembles, namely
ensembles that are defined by degree distributions but are otherwise
completely unstructured. Such ensembles can have good performance
(e.g., we have seen that for the BEC they can achieve capacity) but
``real'' codes typically have additional structure which allows to
optimize various performance metrics. We will now discuss one such
structure which is called {\em spatial coupling}. As we will see, this
structure will allow us to construct capacity-achieving ensembles for
a much broader class of channels and it is nicely grounded in basic
facts from statistical physics.

\subsection{Protographs}
\label{sec:proto}

\begin{figure}
\begin{center}
\includegraphics[width=2in]{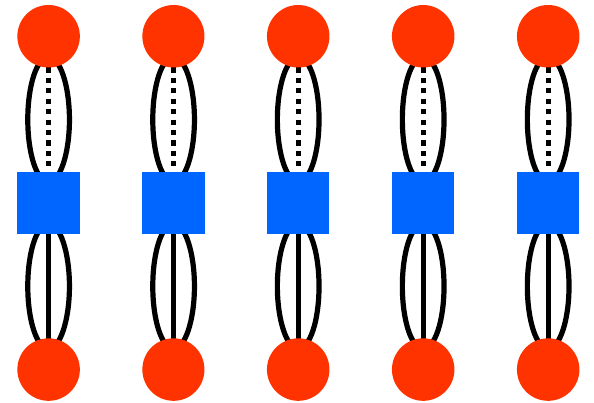}
\hspace{1cm}
\includegraphics[width=2in]{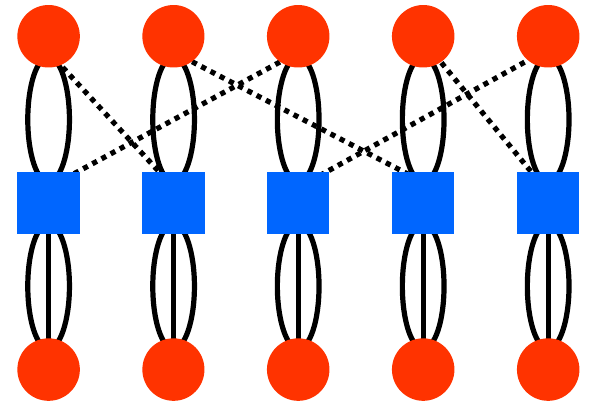}
\end{center}
\caption{In the protograph construction we start with a single
``protograph.'' (see e.g., the left-most graph on the left side).
We then ``lift'' this protograph to a larger graph by taking $M$
copies (in our specific case $M=5$. Finally , we connect the various
copies by taking ``like'' edges (which we call an {\em edge bundle}
and by permuting the edges in the edge bundle via a permutation
picked uniformly at random. One particular edge bundle is shown on
the left  by dotted lines and the result of the permutation is shown
on the right.}
\label{fig:proto_const}
\end{figure}

There are many ways of describing LDPC ensembles and many flavors
of such ensembles. One particularly useful way of describing an
ensemble is in terms of so-called {\em protographs}. This language will
be useful when describing the more complex case of spatially-coupled ensembles.

Protographs were introduced by Thorpe \cite{Thorpe}.  They give a
convenient and compact way of specifying ensembles and the additional
structure they impose is useful in practice.  The creation of a
``real'' graph from protographs is illustrated in
Fig.\ref{fig:proto_const}.  For simplicity, $M=5$ copies are
introduced in Fig.~\ref{fig:proto_const}, but $M$ is typically in
the order of hundreds or thousands.  The edges denoted by dashed
lines in Fig.~\ref{fig:proto_const} are ``edge bundles.'' Such an
edge bundle is a set  of ``like'' edges that connect the same
variable node and the same check node in each protograph.  In a
protograph we connect the $M$ copies by permuting the edges in each
edge bundle by means of a permutation chosen uniformly at random
as shown in right of Fig.~\ref{fig:proto_const}.  Strictly speaking,
the ensemble generated in this way is different from the ensemble
generated by the configuration model, but these models are
asymptotically equivalent in the sense that density evolution as
discussed before gives the correct asymptotic predictions in both
cases.

\subsection{Construction of Spatially Coupled Codes}
Let us now introduced spatially coupled ensembles.  There are as
many flavors and variations of spatially coupled codes as there are
for uncoupled codes.  The exact version we consider here is not so
important since they all behave more or less the same.  Hence, let
us consider two variants that are easy to describe and are typical.
The first is a protograph-based construction whereas the second one
is purely random.

\subsubsection{Protograph construction}
\begin{figure}
\begin{center}
\includegraphics[width=2.0in]{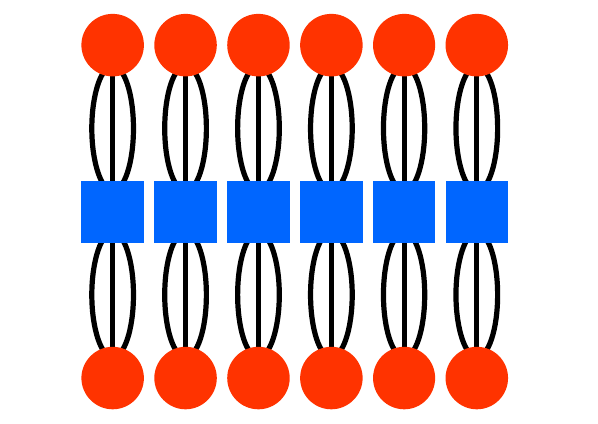}
\includegraphics[width=2.0in]{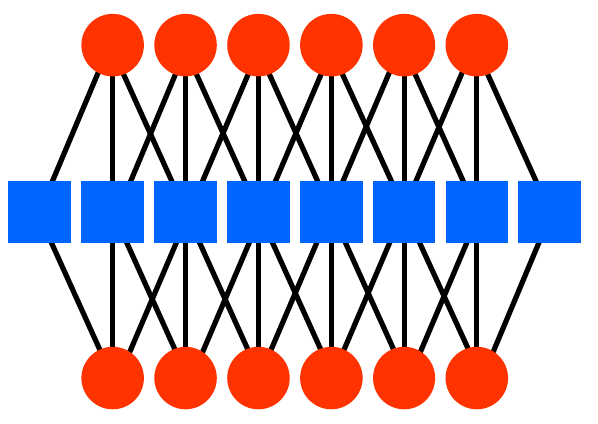}
\end{center}
\caption{Spatially coupled codes (right) constructed from a set of 
$(3,6)$ protographs (left).}
\label{fig:SC_proto}
\end{figure}

In the protograph construction, we start by taking a certain number
of like protographs and placing them next to each other on a line
as shown on the left in Fig.~\ref{fig:SC_proto}. We then ``connect''
{\em neighboring} copies in a regular fashion as shown on the right
in the figure. This gives us a protograph which has a spatial
structure, explaining the origin of the name ``spatially coupled.''
Note that towards the middle of the chain the degree structure of
the graph is exactly the same as the degree structure of the
protograph we started with. Only towards the boundary, due to
boundary effects do we have a different degree structure. Note that
a variable node in the picture is connected to $3$ different
positions. We therefore say that the ``connection width'' is $3$
and we write $w=3$.  At the boundaries, the code has more available
information in the sense that the number of edges are less than the
middle part as shown in Fig.~\ref{fig:SC_proto}.  As we will see,
this boundary condition plays a crucial role.

Note that the right picture in Fig.~\ref{fig:SC_proto} is not the
graph (code) itself yet but just a protograph representing the code.
As mentioned in Sec.~\ref{sec:proto}, to generate the real code
from a given protograph, we need to ``lift'' the graph $M$ times
and then randomly permute edges in the same edge bundle.

Note: Coupled codes constructed in this way from protograph show
an excellent performance and are ideally suited for implementation
by virtue of the additional structure. But they are more difficult
to analyze than the random construction which we discuss below.

\subsubsection{Random construction}

\begin{figure}
\begin{center}
\includegraphics[width=2.0in]{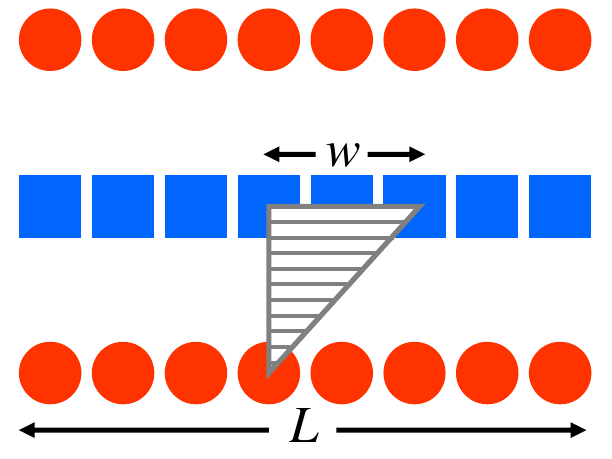}
\end{center}
\caption{Random construction of spatially coupled codes.
Edges are defined randomly in the shaded area.}
\label{fig:SC_random}
\end{figure}
In the random construction we have the same spatial structure for
the nodes but edges connecting neighbors a placed in a more random
fashion. More precisely, we randomly connect check nodes and variable
nodes within a window of size $w$ as shown in Fig.~\ref{fig:SC_random}.
Again, we ensure that the degree distribution away from the boundary
is equal to the degree distribution of the original code.  Note
that when $w=L$, then in fact we impose no spatial constraints on
the connectivity, and we recover the standard uncoupled LDPC ensemble.
This randomly constructed coupled ensemble performs slightly worse
in terms of its finite-length performance but it is easier to analyze
since it has fewer parameters.

\subsection{Why spatial coupling }
\index{Y}
\index{Spatial coupling}
Before we proceed with the theoretical analysis of the spatially
coupled ensembles, let us quickly show that spatially coupled
ensembles behave quite differently from uncoupled ensembles when
we let the degrees tend to infinity. Since the local degree
distribution is the same, this will show that the spatial structure
indeed leads to some interesting behavior.

\subsubsection{Degree dependence of the uncoupled ensembles}

\begin{figure}
\begin{center}
\includegraphics[width=3in]{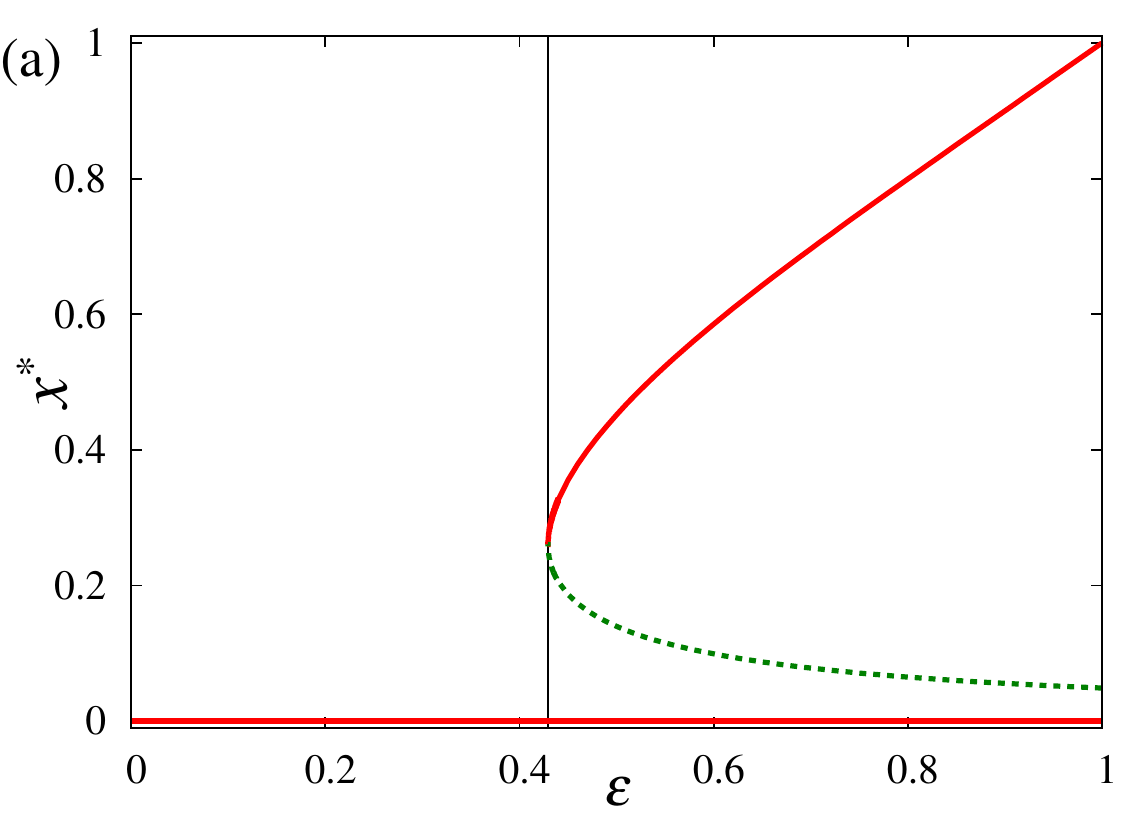}
\includegraphics[width=3in]{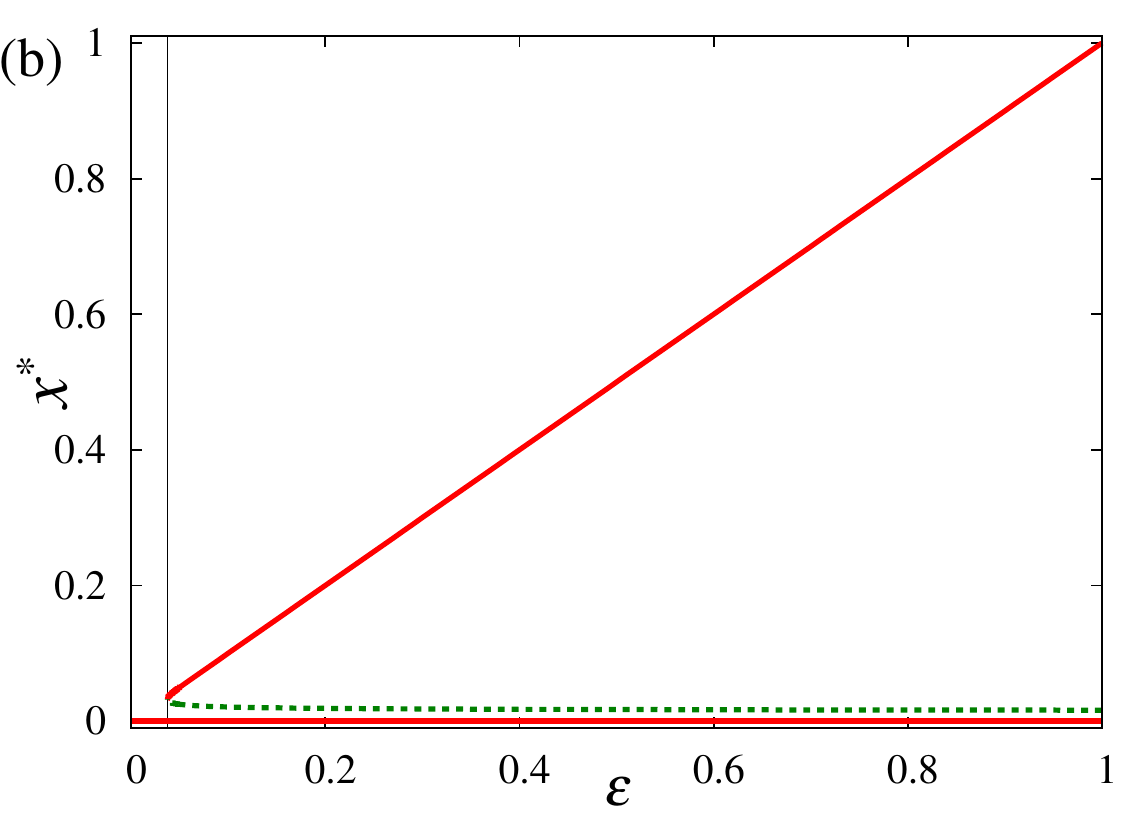}
\end{center}
\caption{Fixed points of uncoupled (a) $(3,6)$ code and (b) $(100,200)$ code.
The vertical lines represent threshold (a) $\epsilon_{\rm BP}\simeq 
0.42944$ and (b) $\epsilon_{\rm BP}\simeq 0.0372964$.
}
\label{fig:EXIT_uncoupled}
\end{figure}

The two pictures in Fig.~\ref{fig:EXIT_uncoupled} show the fixed
points of density evolution for the uncoupled case for (a) the
$(3,6)$ LDPC ensemble and (b) the $(100,200)$ LDPC ensemble. Note
that both have a rate of one-half.  The solid and dashed line
represent stable and unstable fixed points, respectively, and the
vertical lines represent the BP threshold; for (a) we have
$\epsilon_{\rm BP}=0.42944$ and for (b) we have $\epsilon_{\rm
BP}=0.0372964$.  As shown we can see from Fig.~\ref{fig:EXIT_uncoupled},
and as one can show analytically, as we increase the degree the BP
threshold decreases and it reaches $0$ when the degree tends to
infinity. Is this decrease of the threshold due to the fact that
the associated code gets worse as the degrees become larger or is
it the fault of the (suboptimal) BP decoder? A closer look reveals
that the code itself in fact gets better as the degree increases.
But the decoder becomes more and more suboptimal.

\subsubsection{Spatial coupling might help}
\index{Spatial coupling}

Let us now repeat the above experiment with spatially coupled
ensembles.  We will see that they behave very differently.

\begin{figure}
\begin{center}
\includegraphics[width=2in]{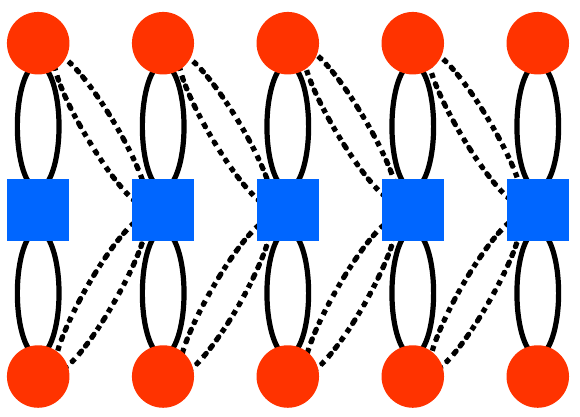}
\end{center}
\caption{A coupled $(2,4)$ ensembles with double edges.}
\label{fig:cyclic_code}
\end{figure}

Consider a coupled ensemble constructed via the protograph approach.
To make the argument particularly simple, assume that all the edges
between factor nodes and variables nodes are in fact double edges,
as shown in Fig.~\ref{fig:cyclic_code}.  E.g., the protograph shown
in this figure therefore represents an $(4, 8)$-regular ensemble.

Consider now the decoder procedure. We want to show that the BP
threshold does not tend to zero for such an ensemble even if we
increase the degrees and let them tend to infinity.

To show this note that we can get a lower bound on the decoding
threshold by ``weakening'' the decoder. We weaken the decoder in
the following way.  Instead of allowing the decoder to use all
available information, assume that when we decode the bits in the
first position we are not allowed to use the information we received
in any of the positions to the right.

This means that for the given example we concentrate only on the
double edges connected to a factor node (denoted by solid lines in
Fig.~\ref{fig:cyclic_code}) and ignore other edges (denoted by dashed
lines).  Note that if we concentrate on the bits in the left-most
position this means that we are decoding a $(2,4)$-regular code,
which is also known as ``cycle code.'' The BP threshold of such a
code is known and e.g. for the BEC it is equal to $\epsilon_{\rm
BP}=1\slash 3$.

Therefore, we know that we can decode the left-most bits using the
BP decoder if we transmit over a BEC with erasure probability not
exceeding $1\slash 3$. Now assume that these positions are known.
We can then remove (the effect of) these bits from the graph.  But
if we do so, what is left looks again exactly like the original
situation except that now the chain is shorter by one. We can
therefore recurse our argument.  In summary, we have just argue
that the BP threshold of this chain is at least one-third.

The punch line is now the following. Exactly the same argument holds
if we increase the degrees and look at the spatially coupled
$(2k,4k)$-regular ensemble, regardless of the value of $k$.  Therefore,
the BP threshold does not tend to zero for coupled ensembles even
if we let the degrees tend to infinity.  This argument only shows
that the threshold is lower bounded by a constant and it does  not
permit to determine the actual threshold. In fact, we will shortly
see that the actual threshold improves as the degree gets larger.

\section{Density Evolution for Coupled Codes}
Let us now get to the analysis of coupled ensembles using the same
method, namely density evolution, which we used in the uncoupled
case.

In the uncoupled case, the variable $x$, which represents the erasure
fraction along an outgoing edge from the variable node, is a scalar
and density evolution tracks the evolution of this scalar as a
function of the iteration number.

For the coupled case the {\em state} is a vector, since variables
at different positions will not experience the same ``environment.''
Recall that at the boundary we have a slightly different degree
distribution and the decoder problem is easier there.  As we will
see, the decoder will be able to decode at the boundary first and
this progress will then propagate towards the interior of the code
along a ``decoding wave.''

Due to this lack of ``homogeneity'' along the spatial dimension we
need a vector $\bm{x}$ to describe the state, where $x_i$ describes
the erasure probability at position $i~(=1,\cdots,L)$.  Recall that
we know the values at the boundary, and hence the erasure probability
at the boundary is $0$.

\begin{figure}
\begin{center}
\includegraphics[width=3in]{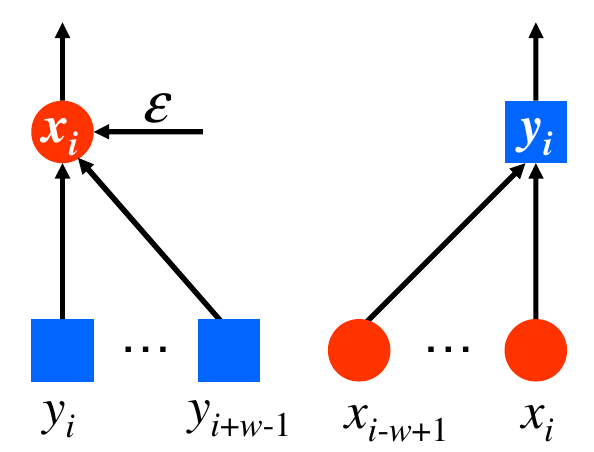}
\end{center}
\caption{Density evolution in spatially coupling code with width $w$.}
\label{fig:SC_XY}
\end{figure}
In the randomly constructed code,
each edge can be connected to positions in a certain range. 
More precisely, consider Fig.~\ref{fig:SC_XY}:
variable nodes assigned $\{x_i\}$ are always connected to position ``to the right'' 
and check nodes assigned $\{y_i\}$ are always connected to variable nodes ``on the left''.
We therefore need to average over the incoming messages from this range,
and the density evolution equations for the coupled ensemble are given by
\begin{align}
x_i&=\epsilon\Big(\frac{1}{w}\sum_{j=0}^{w-1}y_{i+j}\Big)^{d_l-1}\label{eq:DE_couple_x}\\
y_i&=1-\Big(1-\frac{1}{w}\sum_{k=0}^{w-1}x_{i-k}\Big)^{d_r-1},\label{eq:DE_couple_y}
\end{align}
where $i=1,\cdots,L$.  Note that there is an $x_i$ and an $y_i$
value for each position of the chain and the equations for these
values are coupled through the averaging operations.

Combining equations (\ref{eq:DE_couple_x}) and 
(\ref{eq:DE_couple_y}) and adding an index for the iteration number we get
\begin{align}
x_i^{(l)}=\epsilon\Big(1-\frac{1}{w}\sum_{j=0}^{w-1}(1-\frac{1}{w}\sum_{k=0}^{w-1}x_{i+j-k}^{(l-1)})^{d_r-1}\Big)^{d_l-1}.
\label{eq:DE_couple}
\end{align}
To simplify our notation, and also to abstract from the specific
case we are considering, let us define the functions $\{f_i(\cdot)\}$
and $g(\cdot)$,
\begin{align}
f_i&=\Big(1-\frac{1}{w}\sum_{k=0}^{w-1}x_{i-k}\Big)^{d_r-1},
g(\{x_{i\in{ I}(i)}\})=\Big(1-\frac{1}{w}\sum_{j=0}^{w-1}f_{i+j}\Big)^{d_l-1},
\end{align}
where ${I}(i)$ denotes set of indices connected to $i$. In this way we get simple expressions, like
\begin{align}
x_i=\epsilon g(\{x_{i\in{ I}(i)}\}).
\end{align}
We call a vector $\bm{x}=\{x_i\}~(i=1,\cdots,L)$ whose components
are the erasure fractions at the various indices a {\it constellation}.
At all the indices outside the constellation, $i<1$ and $i > L$,
we assume that the corresponding $x_i$ values are 0, i.e., we have
perfect knowledge.  A constellation $\bm{x}$ which when inserted
into the DE equations results in $\bm{x}$ is called a {\em fixed point}
of DE equation.

\begin{figure}
\begin{center}
\includegraphics[width=2.4in]{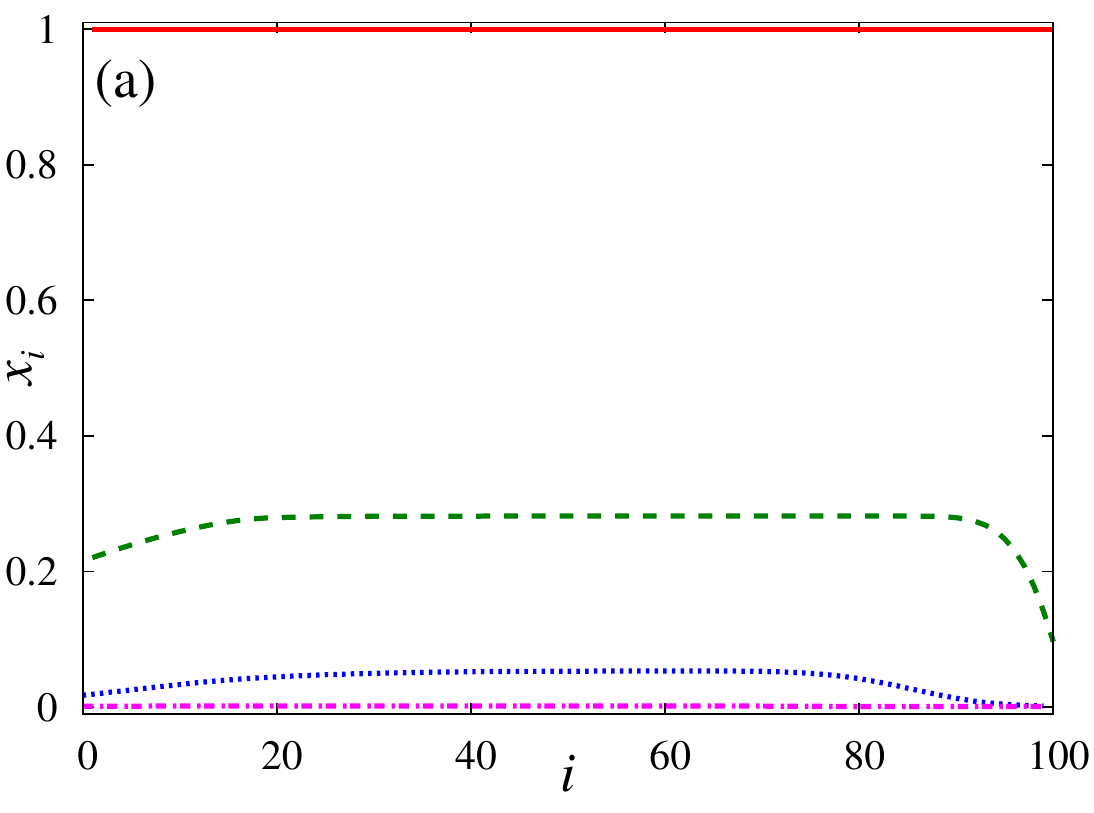}
\includegraphics[width=2.4in]{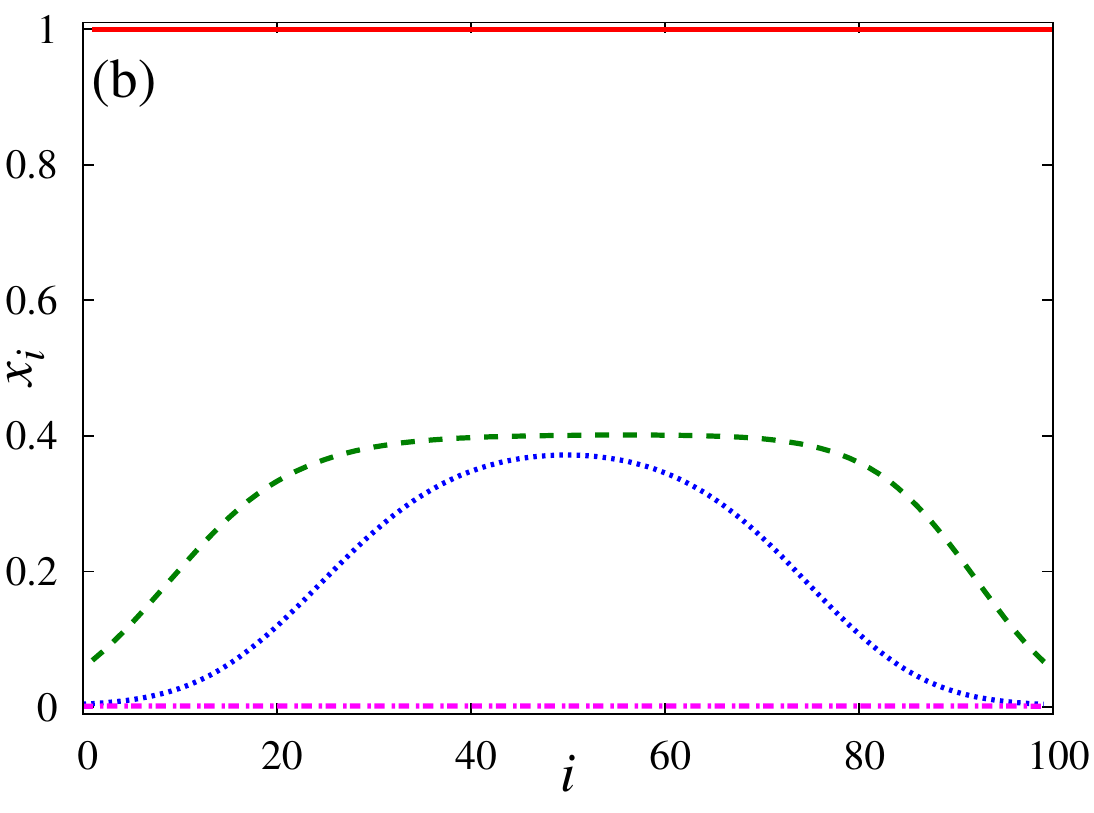}
\includegraphics[width=2.4in]{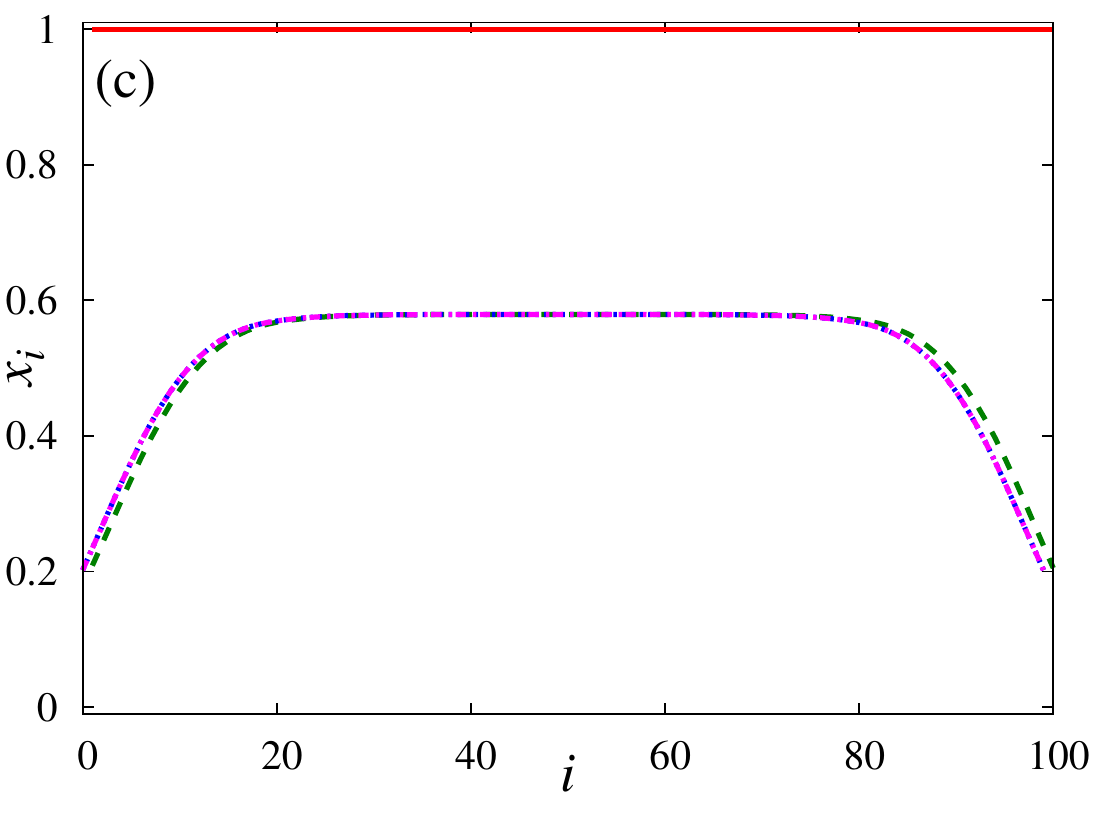}
\end{center}
\caption{Time evolution of $\{x_i\}$ by DE equation for the $(3,6)$
coupled code at (a) $\epsilon=0.3$, (b) $\epsilon=0.48$, and (c)
$\epsilon=0.6$.  The length is $L=100$ and width is $w=20$.  The
constellations evolve in the order of solid line $\to$ dashed line
$\to$ dotted line $\to$ dashed-dotted line.} \label{fig:DE_SC}
\end{figure}

In Fig.~\ref{fig:DE_SC}, the $\epsilon$-dependence of the time
evolution of the DE equation for the coupled ensemble, according
to equation (\ref{eq:DE_couple}), is shown.  Picture (a) corresponds
to $\epsilon=0.3$, (b) corresponds to $\epsilon=0.48$, and (c) is
for $\epsilon=0.6$.  Note that for $\epsilon<\epsilon_{\rm BP}\simeq
0.4294$, DE proceeds in essentially exactly the same way as for the
uncoupled case if we look at the $x_i$ values in the center of the
chain. At the boundary we see somewhat better values due to the
boundary condition.  And as expected, the DE is able to drive the
erasure fraction in each section to zero and BP is successful.

At $\epsilon=0.48$, which is considerably larger than the BP threshold
$\epsilon_{\rm BP}=0.4294$ of the uncoupled ensemble (and close to
the optimal threshold of $0.5$ of the best code and decoding
algorithm), a small ``wave front'' is formed at both boundaries
after a few iterations due to the fact that at the boundaries more
knowledge is available, see Fig.~\ref{fig:DE_SC} (b).  These wave
fronts move towards the center of the coupled code at a constant
speed and by doing so decrease the value of $x_i$ for $i$ located
in the central part of the coupled code until the whole constellation
is decoded. This is the interesting new phenomenon that happens due
to the spatial structure. In other words, due to the spatial
structure, the wave front can smoothly connect the desired fixed
point of $x_i=0$ to the undesired fixed point that is found by the
BP decoder of the uncoupled system and at a constant speed the
undesired fixed point is guided towards the desired one until
decoding is accomplished. As we increase the parameter $\epsilon$
up to a critical threshold, call it $\epsilon_{\rm Area}$, the speed
of the wave is linearly decreased and it reaches the value zero at
$\epsilon_{\rm Area}$.

At $\epsilon$ above $\epsilon_{\rm Area}$, we get a non-trivial
fixed point of DE and decoding is no longer successful.  In the
middle of the chain the $x_i$ values are exactly as large as they
would be for the same $\epsilon$ value in the uncoupled case.  Only
at the boundary do we get somewhat better values because of the
boundary condition.

\subsection{Summary}
In the following sections, we will see that spatially coupled
ensembles can be decoded up to $\epsilon_{\rm Area}$ and this value
is essentially equal to $\epsilon_{\rm MAP}$ of the underlying
ensemble.  This phenomenon is called {\em threshold saturation}.
In order to exactly achieve $\epsilon_{\rm MAP}$, we have to let
the chain length $L$ tend to infinity (this makes decoding harder)
and the interaction width $w$ tend to infinity as well (with $w <<
L$).  But in practice, even for moderate values of $L$ of perhaps
$10$ or $20$ and very small values of $w$, perhaps $2$ or $3$, the
decoding thresholds are already very close to $\epsilon_{\rm MAP}$;
for instance for the BEC, the difference between $\epsilon_{\rm
MAP}$ and $\epsilon_{\rm Area}$ with $w=3$ is only about $10^{-5}$ for
the $(3,6)$-regular ensemble.

\index{MAP}
Note finally, that one can show that the MAP threshold $\epsilon_{\rm
MAP}$ is an increasing function of the degrees and converges to the
Shannon threshold exponentially fast in the degrees.  This is
\index{Shannon}
contrary to the BP threshold $\epsilon_{\rm BP}$ for uncoupled codes
which typically decreases in the degree.

\section{Threshold Saturation}
Let us now discuss why \index{Y} threshold saturation happens and how we can prove the
above assertions.

We will limit our discussion to the simplest case, namely transmission
over the BEC.  Currently there are proofs of the threshold saturation
phenomenon for the following cases; sparse graph codes and transmission
over any BMS channel, any system whose state (for the uncoupled
system) is a scalar or a vector, and compressive sensing.

For the BEC, there are currently three known proof strategies; via
the Maxwell construction, via EXIT charts, and via potential
functions.  These proofs share important features but each also
have their own advantages.

Historically speaking, the proof of threshold saturation via the
Maxwell construction was the first proof that spatially coupled
codes achieving capacity under BP decoding when transmitting over
the BEC.  Later on the same approach led to the proof that spatially
coupled codes universally achieving capacity under BP decoding over
the whole class of BMS channels.  The details of the proof for the
BEC can be found in \cite{ThresholdSaturation} whereas the general
case is described in \cite{ThresholdSaturation_general}.

Recall that we are interested in finding the largest channel parameter
$\epsilon$ so that the DE recursion of the coupled system, when
started with the all-one vector inside the range $[1,L]$, converges
to the all-zero vector.  We denotes this  parameter by $\epsilon_{\rm
Area}$, and called it the area threshold.

\subsection{Proof by Maxwell construction}

The proof by Maxwell construction consists of three parts; show the
existence of a special fixed point of the coupled DE equations
explained in Sec.~\ref{sec:special_fixed_point}, prove that any
such FP must have a channel parameter that is very close to the
area threshold $\epsilon_{\rm Area}$, and, finally, show that for
any channel parameter $\epsilon$ below $\epsilon_{\rm Area}$, the
DE equations converge to the all-zero constellation $\bm{x}=\bm{0}$.

\subsubsection{Definition of area threshold $\epsilon_{\rm Area}$}

\begin{figure}
\begin{center}
\includegraphics[width=3in]{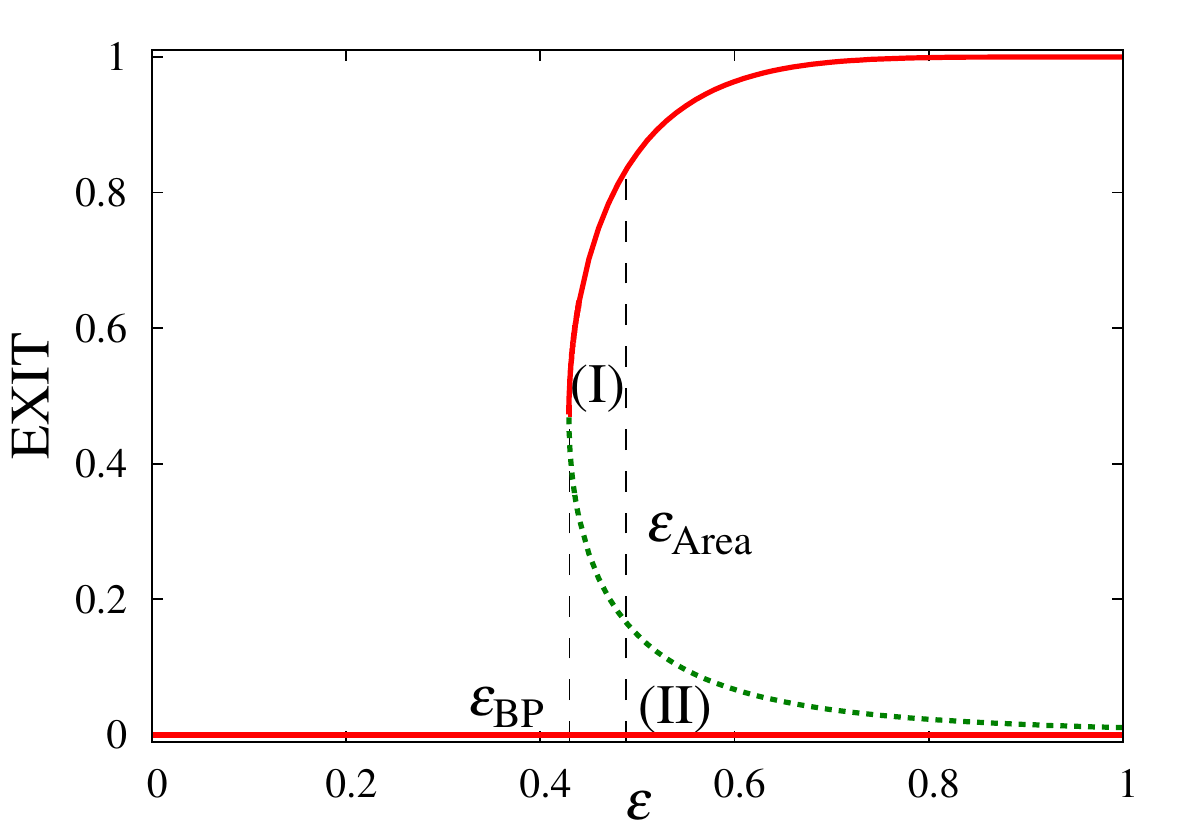}
\end{center}
\caption{Definition of the area threshold $\epsilon_{\rm Area}$.}
\label{fig:def_area}
\end{figure}

Consider Fig.~\ref{fig:def_area}. This figure shows the so-called
EXIT curve for the $(3, 6)$-regular uncoupled ensemble. Recall that
this EXIT curve is the curve that we get if we project the fixed
points of density evolution. The branch plotted as a solid line
corresponds to the stable fixed points, whereas the branch plotted
as dots corresponds to the unstable fixed points. Recall that
equation (\ref{equ:fixedpoint}) gives an explicit description of
these fixed points, i.e., it expresses the channel parameter
$\epsilon$ as an explicit function of the erasure probability $x$
emitted by variable nodes, call this function $\epsilon(x)$.
Explicitly, the EXIT curve is the curve given in parametric form
as $\{x^{d_l}, \epsilon(x))\}_{x=0}^{1}$.

In terms of this EXIT curve the area threshold is defined as follows.
Integrate the area enclosed under the top (stable) branch of the
EXIT curve starting from the right $(\epsilon=1$) until that channel
parameter so that this area is equal to the rate of the code. For
the example shown in Fig.~\ref{fig:def_area} the rate is equal to
$\frac12$ and $\epsilon_{\rm Area}\simeq 0.48818$.

Recall that for this example the BP threshold $\epsilon_{\rm BP}\simeq
0.4299$ so that the area threshold is (considerably) larger than
the BP threshold. This is always the case. It is also easy to see
that the area threshold is always lower than the Shannon threshold
\index{Shannon}
since the EXIT curve is upper bounded by 1; ${\rm EXIT}(\epsilon)\leq
1$ for any $\epsilon\in[0,1]$, and so the area threshold is upper
bounded by $\epsilon_{\rm Area}\leq 1-{\rm rate}$.

By simple explicit calculation, it can be further shown that the
area which is contained  ``inside'' the ``C''-shaped EXIT curve is
also equal to the rate.  This implies that the area (I) and (II)
shown in the Fig.~\ref{fig:def_area} are equal to each other.
Therefore, an equivalent definition of the area threshold is to say
that it is that point where a vertical line makes the two areas to
be of equal size.

Because of the similarity between the definition of area threshold
and Maxwell construction in thermodynamics, this line is called
Maxwell construction of BP EXIT curve \cite{MaxwellConstruction}.
Indeed whereas in the original Maxwell construction the areas
represent work, in the coding context the areas represent information
which on the one hand a genie has to provide to the BP decoder in
order to convert it into a MAP decoder (area (I)) and on the other
side the amount of ``confirmations'' that the BP decoder receives
during the decoding process that proves that the information provided
by the genie is indeed correct. When these two areas are in balance
then the BP decoder can with high probability decode (just like the
MAP decoder could do) and can at the end certify that all the
information provided by the genie is indeed correct.

\subsubsection{Existence of a special fixed point}
\label{sec:special_fixed_point}

\begin{figure}
\begin{center}
\includegraphics[width=2.4in]{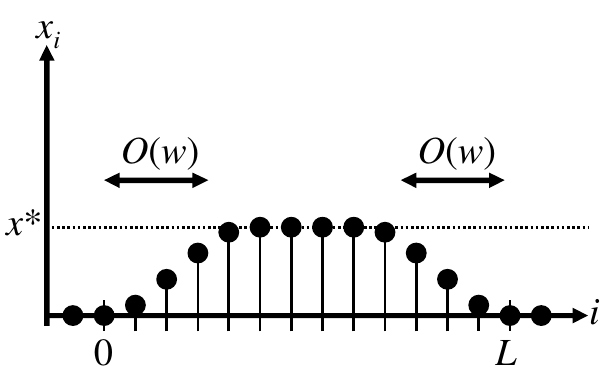}
\includegraphics[width=2.4in]{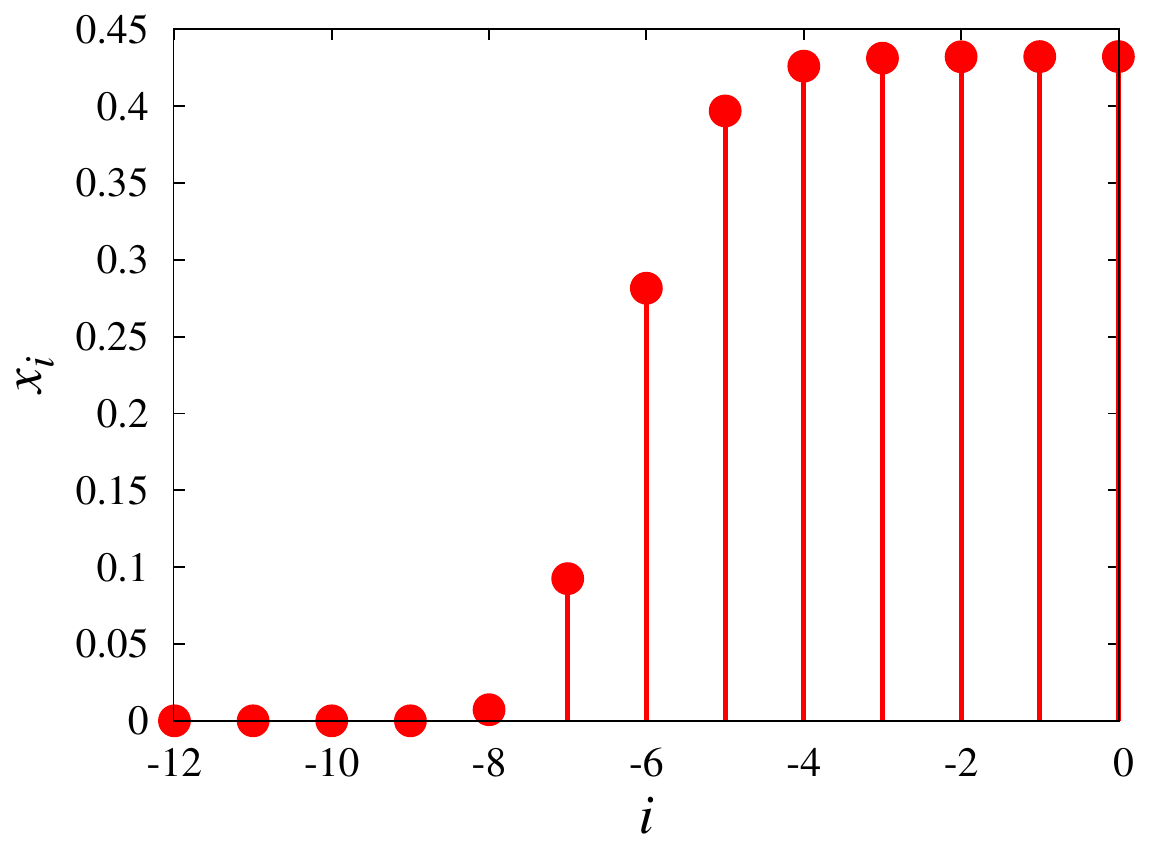}
\end{center}
\caption{(left) Schematic picture of the special fixed point.
(right) One-sided constellation of the special fixed point for $(3,6)$ 
coupled code with $L=6$ and $w=3$.}
\label{fig:fixed_point}
\end{figure}

The special fixed point of $\bm{x}$ that we need is illustrated in
the left of Fig.~\ref{fig:fixed_point}.  What we need is a fixed
point that is unimodal; where $x_i$ is close to $0$ close to the
boundary, and close to $x^*$ in the middle, respectively;
here $x^*$ is the fixed point of DE for the uncoupled system under
the same channel parameter.  Further, the number of positions $i$
whose $x_i$ value is in the range $[\delta,x^*-\delta]$, $\delta>0$,
must be of order $O(w)$.  Note that the two stable fixed points of
DE for the uncoupled system, namely $0$ and $x^*$, are essentially
the lower and upper bounds on $\bm{x}$ and that $\bm{x}$ should
smoothly interpolate between them.

For simplicity, we consider DE for one-side constellations
$(x_{-L},\cdots,x_0)\in[0,1]^{L+1}$ as shown in the right of
Fig.~\ref{fig:fixed_point}, where $L$ is the length of the chain
and $x_{-L}<\cdots<x_0$.  The DE equation for one-side constellations
is obtained from the usual coupled DE equation eq.~(\ref{eq:DE_couple})
by setting $x_i=0$ for $i<-L$ and $x_i=x_0$ for $i>0$.  We define
the average value (entropy) of the one-side constellation as
\begin{align} 
\overline{x}=\frac{1}{L+1}\sum_{i=-L}^0x_i.
\end{align}
We can establish the existence of the fixed point with the desired
properties by the use of Schauder's fixed point theorem, which
states that any continuous mapping $f$ from a convex compact subset
$S$ of a Euclidean space to $S$ itself has a fixed point.  In fact,
when applying the fixed point theorem we do not fix the parameter
$\epsilon$, but this parameter is part of fixed point itself.
Therefore, as a consequence of Schauder's fixed point theorem, after
some proper definition of the fixed point equation we are guaranteed
the existence of a constellation $\bm{x}^*$ with the desired
properties which is a fixed point for some channel parameter
$\epsilon^*$. Although one can establish a priori bounds on the
range of $\epsilon^*$, its exact value is not known. This is the
point of the next step in the proof.

\subsubsection{Saturation}

\begin{figure}
\begin{center}
\includegraphics[width=3in]{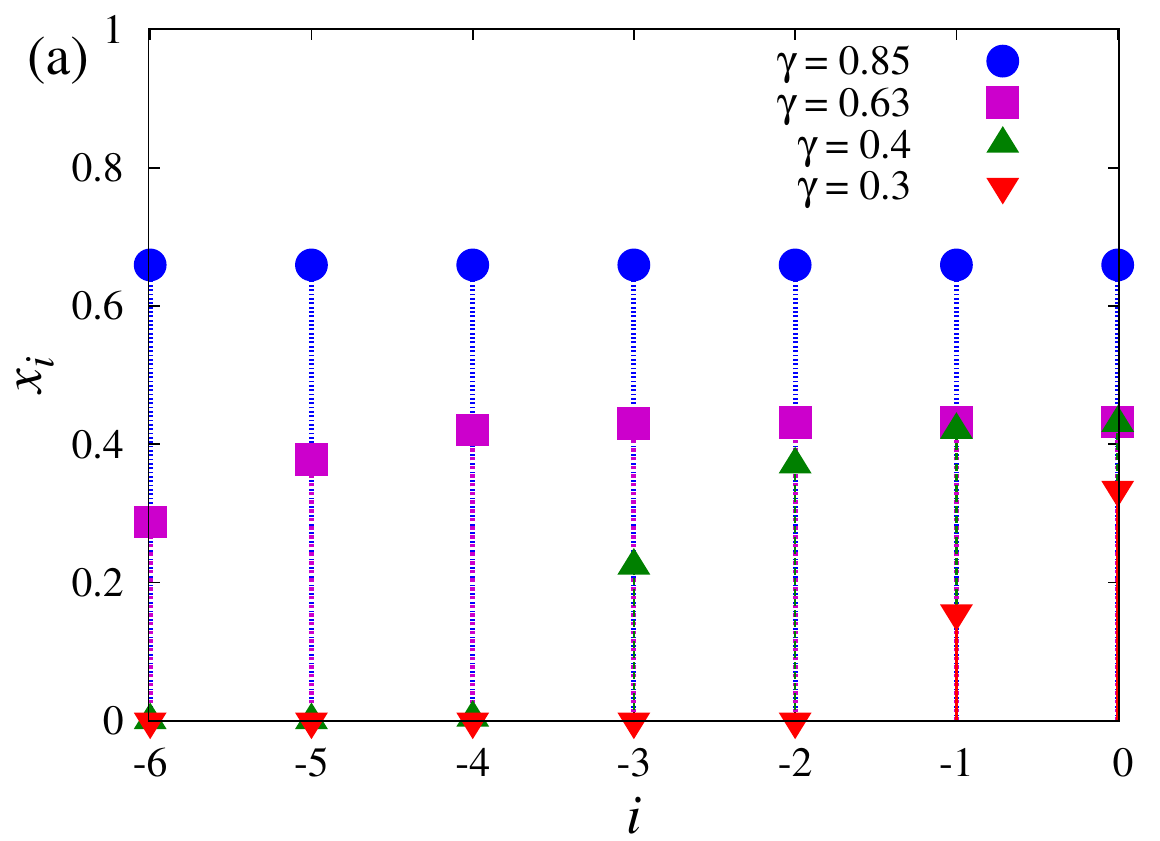}
\includegraphics[width=3in]{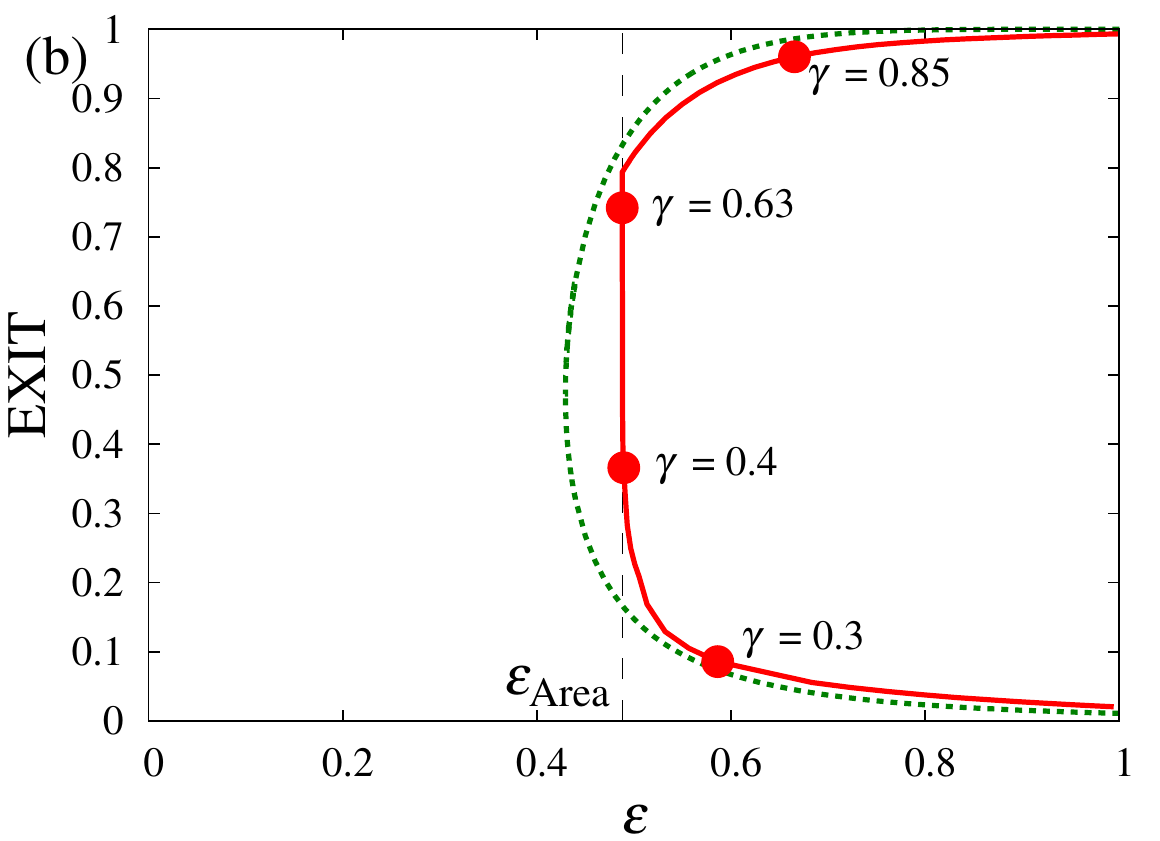}
\end{center}
\caption{(a) Examples of interpolated family for $(3,6)$ coupled code with $L=6$ and $w=2$
and (b) corresponding EXIT curve. Dashed line denotes EXIT curve of uncoupled case.}
\label{fig:interpolation}
\end{figure}
Next, we show that when we have the special fixed point, its channel
parameter must be very close to $\epsilon_{\rm Area}$. The basic
idea is very simple.  Recall our discussion of the EXIT curve for
the uncoupled system. In this case we mentioned that the area
enclosed ``within'' this EXIT curve is equal to the rate of the
code. For the uncoupled case this was the result of a simple explicit
computation since the EXIT curve was known in parametric form and
the integration can be carried out without problems. But there
exists also a more conceptual proof which does not rely on explicit
calculations and which shows that any time you have a smooth EXIT
curve the area it encloses must be equal to the rate of the code.
\index{Y}Why is this true?  It turns out that the EXIT curve can be interpreted
as the derivative of an entropy term with respect to the channel
parameter and so when we integrate, by the fundamental theorem of
calculus, the area is just the difference of this entropy term at
the two end points. This difference can be determined explicitly
and it happens to be equal to the rate of the code. More is true,
assume that instead of have a real EXIT curve, where we recall that
each point corresponds to a fixed point of DE) we have a smooth
curve where every point corresponds to an ``approximate''  fixed
point of density evolution. Here, ``approximate'' means that the
difference of the point and the point we get after one iteration
is small in the appropriate metric. In this case the same conceptual
argument tells us that the area enclosed by this curve is ``close''
to the rate of the code, where the measure of ``closeness'' is
related to how close the points are to being fixed points.

The idea is  hence the following. Given the special fixed
$(\epsilon^*,\bm{x}^*)$ we will construct from it a whole family
of approximate fixed points so that this family gives rise to an
approximate EXIT curve.  The shape of the approximate EXIT curve
is the one shown as a solid curve in Fig.~\ref{fig:interpolation}(b).
In particular, the sharp vertical drop happens exactly at the
parameter $\epsilon^*$ and the whole EXIT curve will look just like
the curve we get from the Maxwell construction. Applying then the
fact that the integral must be equal to the rate of the code will
tell us that the sharp vertical drop must happen exactly at the
area threshold.

But how can we construct from this single special fixed point a
whole family?  Rather than discussing the whole construction let
us only discuss the most interesting part, namely the part corresponding
to the sharp vertical drop. Recall that one of the conditions on
the special fixed point was that in the ``middle'' the fixed point
was essentially flat and had a value essentially equal to what the
uncoupled ensemble would have for this channel parameter.  Further,
towards the boundary the boundary the values had to be essentially
equal to $0$. This means that we insert any number of further
sections in the middle with the appropriate value or any number of
further sections at the boundary with the value $0$ and we will
still have an appropriate fixed point. All of them will be appropriate
fixed points corresponding to the same channel value but their
average value will depend on how wide we make the middle part.  By
changing this width we get points on the vertical line. Since the width
can only be changed in discrete steps but we need a continuous curve 
we also need to interpolate the discrete steps. In addition, in order to get the
points on the top horizontal portion of the EXIT curve we also need to interpolate.
This is shown in  Fig.~\ref{fig:interpolation}(a).

\subsubsection{Convergence}

\begin{figure}
\begin{center}
\includegraphics[width=3in]{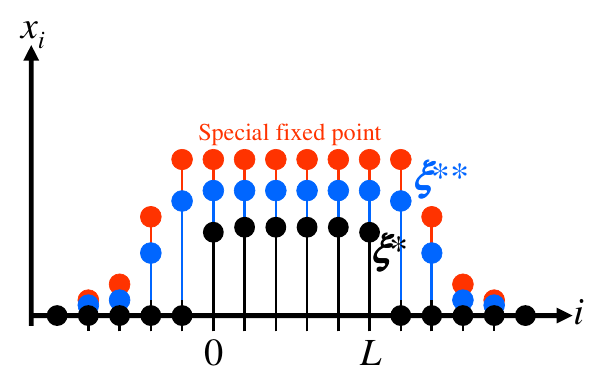}
\end{center}
\caption{Assumed fixed point $\bm{\xi}^*$ at $\epsilon_{\rm
BP}<\epsilon<\epsilon_{\rm Area}$, special fixed point, and a fixed
point $\bm{\xi}^{**}$ obtained by applying DE for the special fixed
point at $\epsilon$.} \label{fig:convergence} \end{figure}

We now get to the last part of the argument. By now we have established
that such a special fixed point can only exist if its channel
parameter is very close to the area threshold. We will now argue
that if we start DE with a channel value below this area threshold
that it must converge to the all-zero constellation.

To see this, we consider the following experiment.  We apply DE at
$\epsilon_{\rm BP}<\epsilon<\epsilon_{\rm Area}$ to a constellation
of size length $L$ whose initial condition is the all-one vector
inside the constellation and 0 outside.  DE produces a sequence of
monotonically decreasing (point-wise) constellations which are
bounded by 0 (again point-wise) from below.  We denoted the fixed
point by $\bm{\xi}^*$ and assume that $\bm{\xi}^*$ is non-trivial,
i.e., is not all-zero as shown in Fig.~\ref{fig:convergence}.  It
is clear that at each point in the constellation, the value of the
fixed point is no larger than the fixed point we would get for the
uncoupled case at $\epsilon$ since at the boundary the decoder has
access to additional information.

Now let us compare this fixed point to our special fixed where we
pick the length for this special fixed point sufficiently large so
that this special fixed point dominates $\bm{\xi}^*$ everywhere
point-wise. Note that $\bm{\xi}^*$ is a fixed point for the parameter
$\epsilon$ but the special fixed point is for the parameter
$\epsilon_{\rm Area}$ and $\epsilon<\epsilon_{\rm Area}$. So if we
now apply DE to the special fixed point but with the parameter
$\epsilon$ then the special fixed point must be decreasing strictly
point-wise and it must in fact converge to the all-zero constellation
since otherwise we would get another non-trivial fixed point which
would again fulfill all the requirements of a special fixed point
(this needs some arguments to prove this) and we know that the only
channel parameter for which such a special fixed point exists is
very close to $\epsilon_{\rm Area}$, a contradiction. But since our putative fixed point
$\bm{\xi}^*$ is dominated by our special fixed point and the special fixed point collapses
to the all-zero constellation it must in fact be true that
$\bm{\xi}^*$ is also the all-zero constellation.

\subsection{Proof by EXIT charts}

\subsubsection{EXIT charts}

EXIT charts were introduced by S. ten Brink as a convenient way of
visualizing DE \cite{Brink}.  For transmission over the BEC, the
EXIT chart method is equivalent to DE.  EXIT charts and EXIT curves
which we have already introduced are quite different despite their
similar name.  The reason both objects have the word ``EXIT'' in
there is that in both cases we measure the same thing (namely if
the ``other'' bits in a code are able to determine the bit we are
considering via the code constraints), but for EXIT charts we make
local measurements, whereas for EXIT curves we measure the performance
of the whole code.

\begin{figure}
\begin{center}
\includegraphics[width=2.4in]{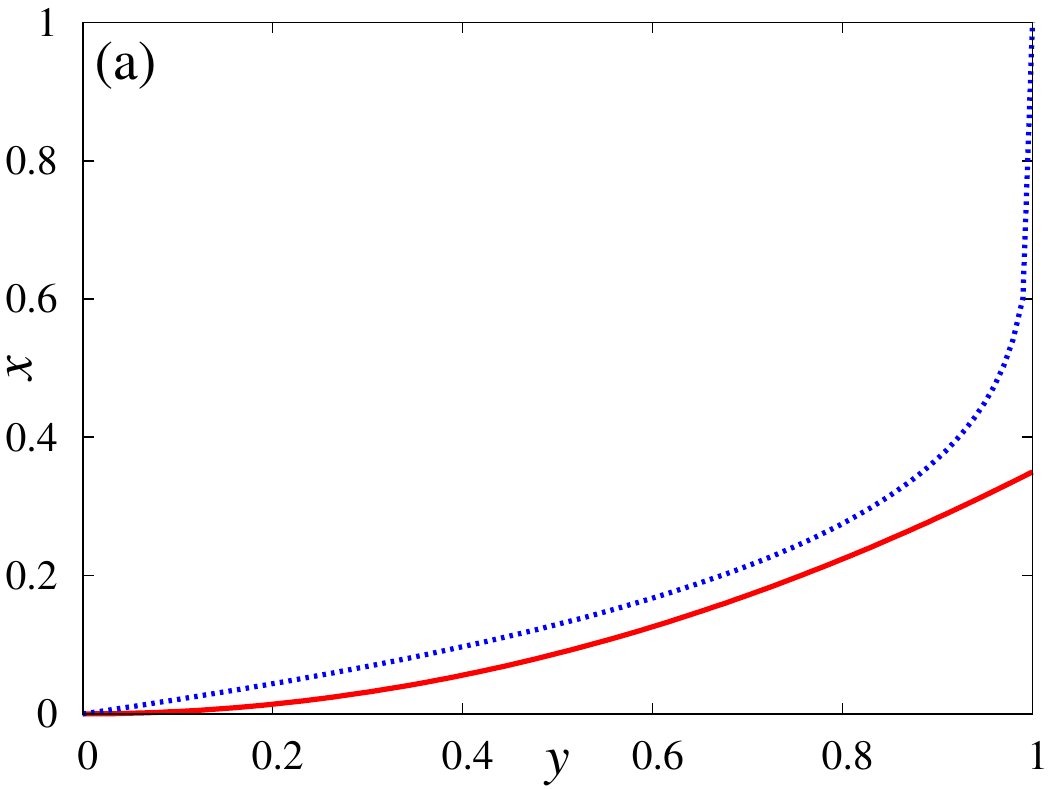}
\includegraphics[width=2.4in]{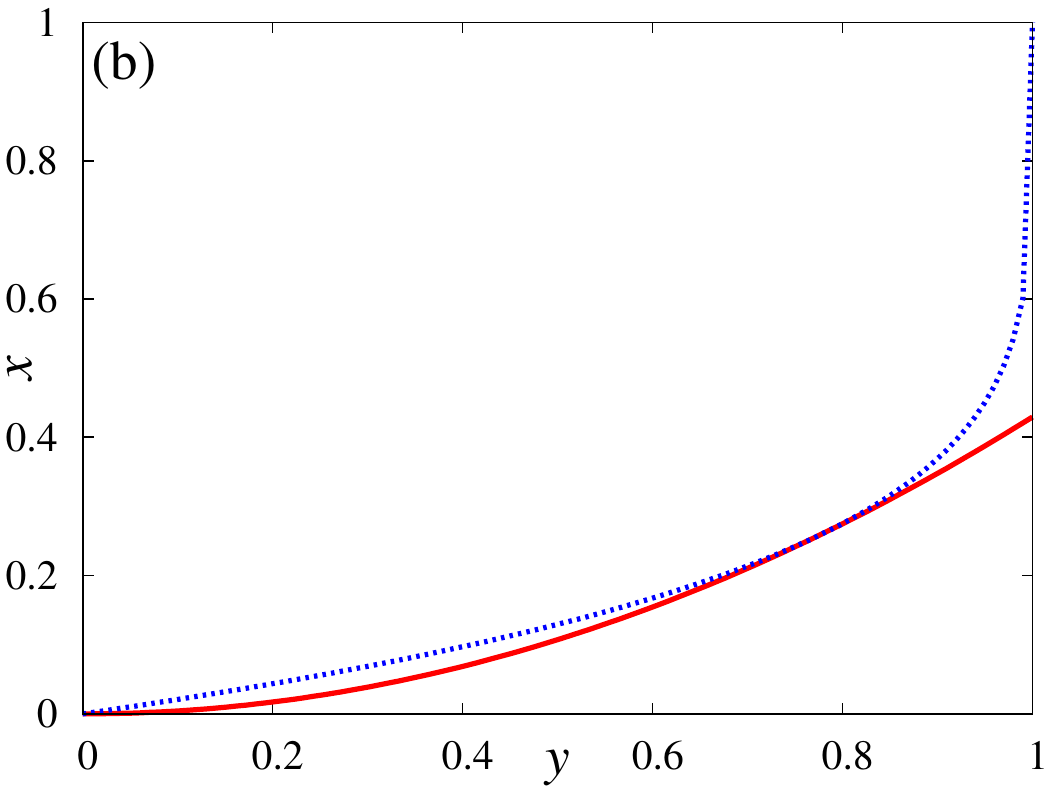}
\includegraphics[width=2.4in]{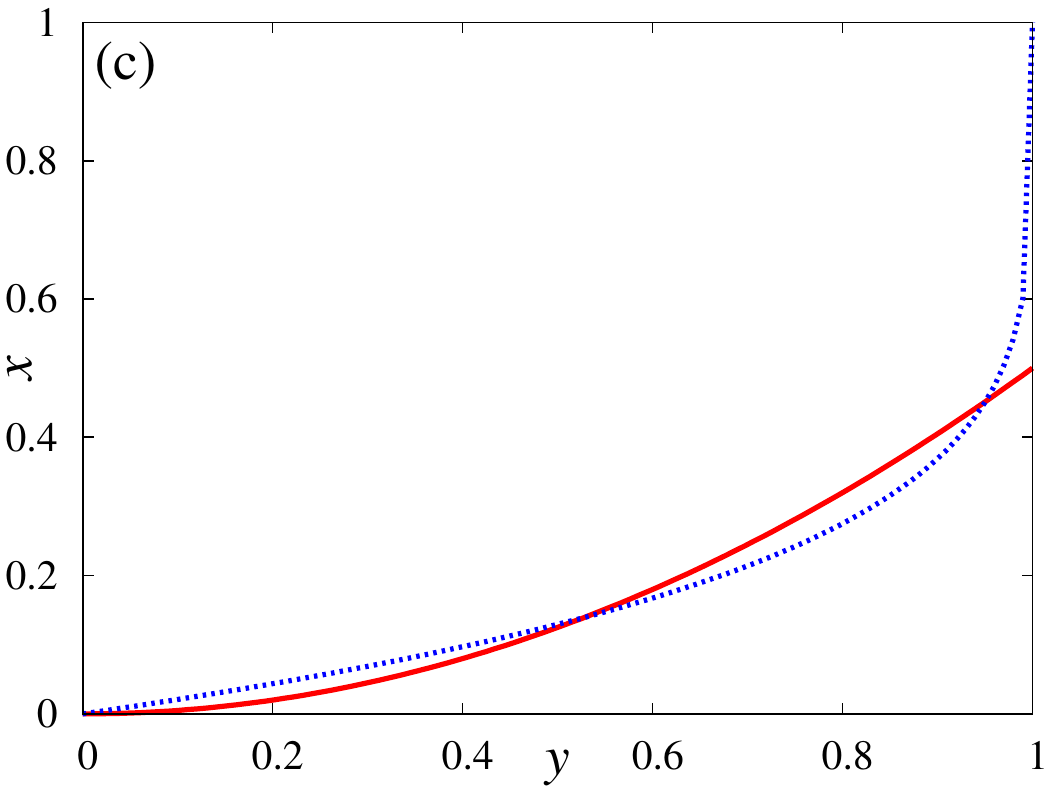}
\end{center}
\caption{EXIT charts of $(3,6)$ uncoupled LDPC
for (a) $\epsilon=0.35$, (b) $\epsilon=\epsilon_{\rm BP}\simeq 0.4294$,
and (c) $\epsilon=0.5$.}
\label{fig:EXIT_chart_uncouple}
\end{figure}

An EXIT chart consists of two curves.  One curve corresponds to the
message-passing rules at the variable nodes and the other one to the
message passing rules at the check nodes. In addition, it is
customary, and it is convenient, that we plot one curve with its
input on horizontal axis and its output on the vertical axis and
the other curve is the plot with its output on horizontal axis and
its input on the vertical axis.  Fig.~\ref{fig:EXIT_chart_uncouple}
shows the EXIT charts of uncoupled $(d_l,d_r)$-LDPC for $d_l=3$ and
$d_r = 6$ at three channel parameters, where two curves are given
by
\begin{align}
x &= \epsilon y^{d_l-1}\\
y &= 1-(1-x)^{d_r-1}.
\end{align}
On this EXIT chart, the DE trajectory can be regarded as a staircase
pattern bound by these two curves.  The DE points converge to zero
if and only if the two curves do not cross
(Fig.~\ref{fig:EXIT_chart_uncouple}(a)).  The threshold for the
uncoupled case is given by the channel parameter so that the two
EXIT curves just touch but do not cross
(Fig.~\ref{fig:EXIT_chart_uncouple}(b)).  At $\epsilon>\epsilon_{\rm
BP}$, two curves touch at three points
(Fig.~\ref{fig:EXIT_chart_uncouple}(c)).

\subsubsection{Proof by EXIT charts for coupled code}

In the coupled systems, the criterion of the threshold, namely that
the two curves may touch but are not allowed to cross, is relaxed.
To determine the threshold for the couple system, the two EXIT
curves are now allowed to cross but not by too much, and indeed,
the threshold relates to a balance of areas enclosed by two curves.
One can show that the condition for the threshold is exactly the
same as the matching of areas condition which we have seen in the
Maxwell construction.

The first step of the proof consists of considering an appropriately chosen
continuous version of the constellation.
For the random coupled ensemble, we have introduced the window $w$,
within which the random connections are generated. 
This discrete system is difficult to analyze.
Instead, one can consider the limit when $w$ goes to infinity,
of course, the length of the chain has to go to infinity as well. 
If we increase the length $w$ and
scale the length of the code by the same proportion 
then in the limit we can treat the constellation as a
continuous curve rather than a set of spikes. 
DE equation for this continuous constellation is 
given by integrating over a window
instead of taking discrete sums.

\begin{figure}
\begin{center}
\includegraphics[width=2.3in]{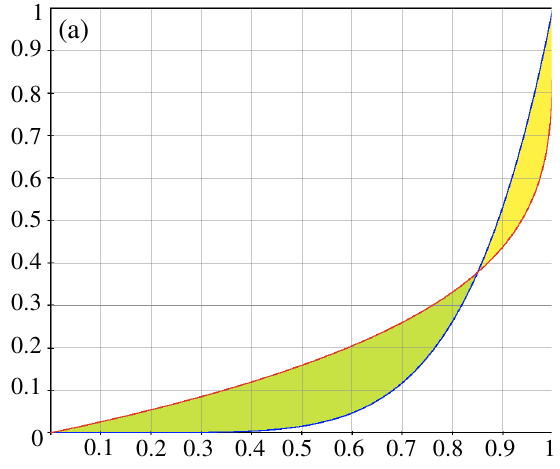}
\hspace{0.1in}
\includegraphics[width=2.3in]{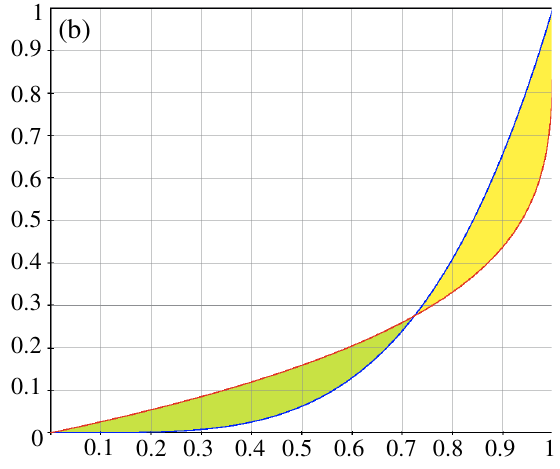}
\hspace{0.1in}
\includegraphics[width=2.3in]{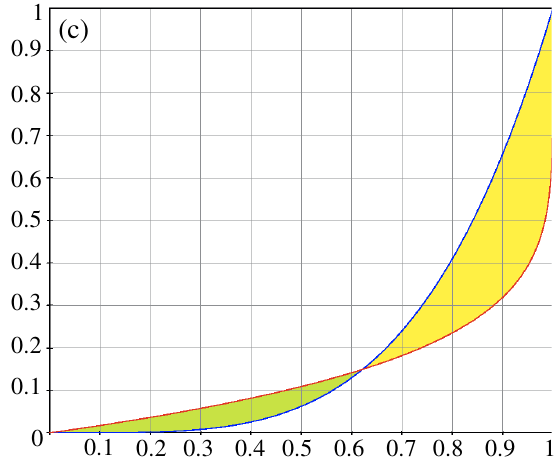}
\end{center}
\caption{EXIT charts of a coupled system for (a) $\epsilon<\epsilon_{\rm 
Area}$, (b) $\epsilon=\epsilon_{\rm Area}$ and (c) 
$\epsilon>\epsilon_{\rm Area}$.
These figures are quoted from lecture materials \cite{OnlineMaterials}.}
\label{fig:EXIT_coupled}
\end{figure}

The next step is to analyze this continuous system. 
It is convenient to think of systems of infinite length, i.e., 
the horizontal axis extends from $-\infty$ to $-\infty$. 
Further, instead of consider a two-sides constellation,
we consider a one-sided constellation,
i.e., we only focus on the ``left'' part of the constellation
from $-\infty$ to $0$.
For the continuous system, 
it is proved that one has three different scenarios
depending on the balance of the areas in the EXIT chart
picture of the uncoupled system. 
We omit here the trivial case where the curves do not overlap at all
since in this case it is easy to show that we will decode.

Consider first the scenario where the channel parameter is below 
$\epsilon_{\rm Area}$, but above the $\epsilon_{\rm BP}$
of the uncoupled ensemble (Fig.\ref{fig:EXIT_coupled}(a)). 
In this case, the curves do overlap but only little and the area on the left (green)
is larger than the area on the right (yellow). 
One can show that there
does not exist a fixed point of DE but there exists a one-sided constellation $\bm{x}$.
If we apply DE to this constellation, 
we get the same $\bm{x}$ but the position is shifted to the right.
Given that our one-sided constellation represents the
left part of an actual constellation, 
saying that the wave is propagating to the right means that the decoder is
working and in each step decodes a further part of the constellation. 
The shift which we see in each
iteration corresponds to the decoding speed and so tells us how many iterations we will need. 
To summarize, 
below the area threshold we get a decoding wave which moves to the right,
that means the working of the decoder.

Assume next that the areas are exactly in balance, 
this means that we are transmitting exactly at $\epsilon_{\rm Area}$ (Fig.~\ref{fig:EXIT_coupled}(b)).
In this case one can prove that the continuous version of DE has a fixed 
point. This fixed point 
can be regarded as a stationary wave or a wave with zero speed.

Finally, consider the case where we are transmitting above 
$\epsilon_{\rm Area}$ (Fig.~\ref{fig:EXIT_coupled}(c)). 
The curves overlap so much so that the area on the left is smaller than the area on the
right.
For this case one can then show that there does not exist a non-trivial 
fixed point but only a continuous
constellation $\bm{x}$, so that after one round of DE we get the same constellation back but shifted to the left.
This means that the decoder does not work.

In a final step one needs to reconnect the continuous system to the actual discrete system and show that if
the $w$ is not too small then the behavior of the discrete system is well predicted by the behavior of the
continuous system \cite{WaveLikeSolution}.

\subsection{Proof by Potential Functions}

\subsubsection{Potential Functions}

\begin{figure}
\begin{center}
\includegraphics[width=3in]{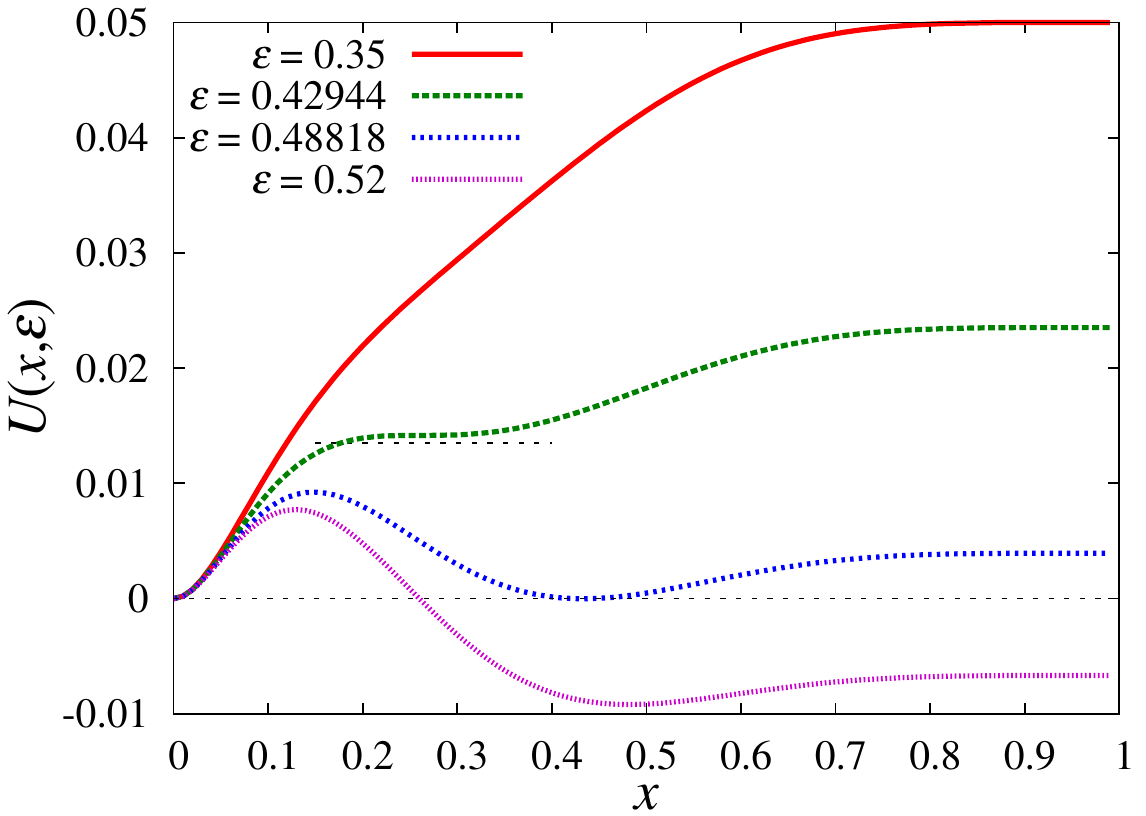}
\end{center}
\caption{Potential function of $(3,6)$ uncoupled LDPC.}
\label{fig:3-6_potential}
\end{figure}

The potential function for uncoupled LDPC code is defined as
\begin{align}
\nonumber
U(x,\epsilon)&=\int_0^x(z-f(g(z);\epsilon))g^\prime(z)dz\\
&=g(x)-G(x)-F(g(x);\epsilon),
\end{align}
where $f(g(z);\epsilon)=\epsilon\lambda(1-\rho(1-z))$,
$g(z)=1-\rho(1-z)$, $F(x;\epsilon)=\int_0^g(z;\epsilon)dz$,
and $G(x)=\int_0^xg(z)dz$.
The functions $\lambda(\cdot)$ and $\rho(\cdot)$
are node-perspective degree distributions.
The potential function corresponds to Bethe free energy.\index{Bethe free energy}
Fig.~\ref{fig:3-6_potential} shows the $x$-dependence of the potential function
for $(3,6)$ uncoupled LDPC, where $\lambda(x)=x^2$ and $\rho(x)=x^5$.
When $\epsilon$ is smaller than $\epsilon_{\rm BP}\simeq 0.42944$,
the potential function is an increasing function for all $x\geq 0$.
At $\epsilon=\epsilon_{\rm BP}$,
the potential function have zero gradient at a certain $x$,
and at $\epsilon>\epsilon_{\rm BP}$,
a local minimum appears and its value touches to the line $0$ at 
$\epsilon_{\rm Area}\simeq 0.48818$.
This means that 
the state $x=0$ is always exists as the unique solution
when $\epsilon <\epsilon_{\rm BP}$.
At $\epsilon>\epsilon_{\rm BP}$,
another locally stable solution appears at $x>0$,
although the solution $x=0$ is the globally stable state.
At $\epsilon>\epsilon_{\rm Area}$,
the solution $x=0$ is no longer a globally stable solution.
In physical terminology,
area threshold and BP threshold
correspond to the first transition point and spinodal point,
respectively.

\subsubsection{Potential Functions for coupled system}

The potential function of the coupled system, which is defined for a 
vector constellation $\bm{x}$,
is introduced in analogy to the uncoupled system.
We consider the potential function for one-sided constellation;
$x_i=0$ when $i$ is not in $[-L,i_0]$
and the value of $x_i$ increases with $i$ up to
$i_0=\lfloor (w-1)\slash 2\rfloor$,
given by \cite{PotentialFunction};
\begin{align}
\nonumber
U(\bm{x};\epsilon)&=\int_{ C}\bm{g}^\prime(\bm{z})(\bm{z}-\bm{A}^{\rm 
T}\bm{f}(\bm{Ag}(\bm{z};\epsilon)))\cdot d\bm{z}\\
&=\bm{g}(\bm{x})^{\rm T}\bm{x}-G(\bm{x})-F(\bm{Ag}(\bm{x});\epsilon),
\end{align}
where $\bm{g}^\prime(\bm{x})={\rm diag}([g^\prime(x_i)])$, 
$[\bm{f}(\bm{x};\epsilon)]_i=f(x_i;\epsilon)$,
$[\bm{g}(\bm{x})]_i=g(x_i)$,
$G(\bm{x})=\int_{ C}\bm{g}(\bm{z})\cdot d\bm{z}=\sum_iG(x_i)$
and $F(\bm{x};\epsilon)=\int_{ C}\bm{f}(\bm{z};\epsilon)\cdot d\bm{z}=\sum_iF(x_i;\epsilon)$.
The matrix $\bm{A}$ is a $(L+3w+i_0+1)\times (L+3w+i_0+1)$ matrix given by
\begin{align}
\bm{A}=\frac{1}{w}\left[
\begin{array}{ccccccc}
1 & 1 & \cdots & 1 & 0 & \cdots & 0 \\
0 & 1 & 1 & \cdots & 1 & \ddots & \vdots \\
\vdots & \ddots & \ddots & \ddots & \ddots & \ddots & 0 \\
0 & \cdots & 0 & 1 & 1 & \cdots & 1 \\
0 & 0 & \cdots & 0 & 1 & \ddots & 1 \\
0 & 0 & \cdots & 0 & 0 & 1 & \vdots \\
0 & 0 & \cdots & 0 & 0 & 0 & 1 
\end{array}
\right],
\end{align}
which gives matrix notation of the DE equation eq.(\ref{eq:DE_couple}).

We set $K_{f,g}=||g^\prime||_{\infty}+||g^\prime||_\infty^2||f^\prime||_\infty+||g^{\prime\prime}||_{\infty}$,
where $||h||_\infty=\sup_{x\in[0,1]}|h(x)|$ for functions 
$h:[0,1]\to\mathbb{R}$ \cite{PotentialFunction}.
Using the potential function,
it can be shown that when $\epsilon<\epsilon_{\rm Area}$ and 
$w>K_{f,g}\slash\Delta E(\epsilon)$,
coupled potential of a non-zero vector decreases by the shifting.
It implies that the value of constellation 
must be reduced by recursion, and the fixed point is the zero vector.
The outline is as follows.
Let us assume that $\bm{x}\neq \bm{0}$ is the unique fixed point of
the one-sided DE equation.
We introduce a right-shifted constellation from $\bm{x}$
as $\bm{Sx}$,
where $[\bm{Sx}]_1=0$ and $[\bm{Sx}]_i=x_{i-1}$ for $i\geq 2$.
The difference of the potential functions between $\bm{Sx}$ and $\bm{x}$ 
is given by
\begin{align}
U(\bm{Sx};\epsilon)-U(\bm{x};\epsilon)=-U(x_{i_0};\epsilon).
\end{align}
Meanwhile, Taylor expansion of $U(\bm{Sx};\epsilon)$
around $\bm{x}$ gives
\begin{align}
\nonumber
U^\prime(\bm{x};\epsilon)\cdot(\bm{Sx}-\bm{x})&\leq 
(U(\bm{Sx};\epsilon)-U(\bm{x};\epsilon))+\frac{K_{f,g}}{w}\\
&<-U(x_{i_0};\epsilon)+\Delta E(\epsilon)\leq 0,
\label{eq:U_Taylor}
\end{align}
where we exploit $w>K_{f,g}\slash\Delta E(\epsilon)$.

When all components of $\bm{Sx}-\bm{x}$
are non-positive, at least one component of $U^\prime(\bm{x};\epsilon)$
should be positive
to satisfy eq.(\ref{eq:U_Taylor}).
The derivative is given by 
$[U^\prime(\bm{x};\epsilon)]_i=g^\prime(x_i)[\bm{x}-\bm{A}^{\rm T}f(\bm{A}g(\bm{x};\epsilon))]_i$
and $g^\prime (x)\geq 0$, and hence
$[\bm{A}^{\rm T}f(\bm{A}g(\bm{x};\epsilon))]<x_i$ should hold. 
It means that one more iteration reduce
the value of the $i$-th component.
This gives a contradiction and means that the fixed point of the 
one-sided constellation is only $\bm{x}=0$
at $w>K_{f,g}\slash\Delta E(\epsilon)$. 

\subsection{Summary}
\index{MAP}
We have shown three proofs of the threshold saturation phenomena.
These three different criteria each have their own advantage. The
first proof makes not only determines the threshold of the coupled
system but makes it clear that this is equal to the MAP threshold.
The EXIT chart approach is convenient for people who are already
familiar to EXIT charts for uncoupled systems and this criterion
is very easy to apply. Finally, the potential function approach
leads to the currently simplest proof of the threshold saturation
phenomenon.

In more detail, let us summarize some of the main points of these
three proofs.  In the proof by the Maxwell construction, at the
area threshold, a special fixed point exists, which has long tails,
quick transition, and large flat part.  This special fixed point
cannot exist below the area threshold (and neither can it exists
at larger channel parameters).  This proof has a connections to
problems in statistical physics and the picture is exactly the same
if we consider transmission over general BMS channels, although the
proofs are more complicated.  Further, it strongly suggests that
the area threshold is also the MAP threshold of the underlying
ensemble.  Indeed, that this is true has recently been shown
\cite{GMU13}.

In the EXIT chart approach, at the area threshold, a stationary
wave exists. Below the area threshold, a propagating wave, traveling
at a constant speed, shows that decoding will be successful.  This
approach, in particular, EXIT charts and the matching condition are
frequently used to analyze systems which have a one-dimensional
state and they are often use to approximately model more general
systems (e.g., Gaussian approximation). For any such system, if we
replace the matching condition with the area balance condition then
we get the equivalent criterion for coupled systems. If the original
state is one dimensional then this criterion is exact, otherwise
it is an approximation in the same way as the matching condition
for EXIT charts is an approximation for uncoupled systems.

Finally, in the potential function approach, at the area threshold,
the potential function has zero gradient.  Below the area threshold,
potential energy is strictly decreasing implying convergence to
perfect decoding.  It leads to the currently simplest known proof
for one dimensional systems.  It can be extended to systems whose
state is no longer a scalar but a vector and even to infinite-dimensional
systems, e.g., general BMS channels.




\end{document}